\documentclass[acmsmall]{acmart}
\usepackage[T1]{fontenc}
\usepackage{enumitem}
\usepackage{amsmath,amsfonts}
\usepackage{algorithmic}
\usepackage{array}
\usepackage{subcaption}
\usepackage{textcomp}
\usepackage{stfloats}
\usepackage{url}
\usepackage{verbatim}
\usepackage{graphicx}
\usepackage{balance}
\usepackage{adjustbox}
\usepackage{booktabs}
\usepackage{multirow}
\usepackage{threeparttable}
\usepackage{arydshln}
\usepackage{tcolorbox}
\usepackage{soul}
\usepackage{fontawesome}
\usepackage{makecell}
\usepackage{natbib}
\usepackage{colortbl}
\usepackage{siunitx}
\usepackage{microtype}
\definecolor{mygray}{gray}{.9}
\usepackage{tikz}
\usepackage[T1]{fontenc}
\usepackage{aecompl}
\usepackage[normalem]{ulem}
\usepackage{hyperref}
\usepackage{graphicx}
\usepackage{subcaption}

\AtBeginDocument{
  \providecommand\BibTeX{{
    \normalfont B\kern-0.5em{\scshape i\kern-0.25em b}\kern-0.8em\TeX}}}

\acmJournal{JACM}
\acmArticle{111}
\AtBeginDocument{
  \providecommand\BibTeX{{
    \normalfont B\kern-0.5em{\scshape i\kern-0.25em b}\kern-0.8em\TeX}}}

\begin{document}

\title{On the Evaluation of Large Language Models in Multilingual Vulnerability Repair}

\author{Dong Wang}
\affiliation{
  \institution{College of Intelligence and 
  Computing, Tianjin University}
  \city{Tianjin}
  \country{China}
}
\email{dong_w@tju.edu.cn}

\author{Junji Yu}
\affiliation{%
  \institution{College of Intelligence and 
  Computing, Tianjin University}
  \city{Tianjin}
  \country{China}
}
\email{junjiyu@tju.edu.cn}

\author{Honglin Shu}
\affiliation{%
  \institution{Kyushu University}
  \city{Fukuoka}
  \country{Japan}
  }
\email{shu.honglin.167@s.kyushu-u.ac.jp}

\author{Michael Fu}
\orcid{0000-0001-7211-3491}
\affiliation{
  \institution{The University of Melbourne}
  \city{Melbourne}
  \country{Australia}
}
\email{michaelfu1998@gmail.com}

\author{Chakkrit Tantithamthavorn}
\affiliation{%
  \institution{Information Technology, Monash University}
  \city{Clayton}
  \country{Australia}
}
\email{chakkrit@monash.edu}

\author{Yasutaka Kamei}
\affiliation{%
  \institution{Kyushu University}
  \city{Fukuoka}
  \country{Japan}
}
\email{kamei@ait.kyushu-u.ac.jp}

\author{Junjie Chen}
\authornote{Corresponding Author}
\affiliation{
  \institution{College of Intelligence and 
  Computing, Tianjin University}
  \city{Tianjin}
  \country{China}
}
\email{junjiechen@tju.edu.cn}

\renewcommand{\shortauthors}{Wang et al.}

\begin{abstract}
Various Deep Learning-based approaches with pre-trained language models have been proposed for automatically repairing software vulnerabilities.
However, these approaches are limited to a specific programming language (C/C++).
Recent advances in large language models (LLMs) offer language-agnostic capabilities and strong semantic understanding, exhibiting potential to overcome multilingual vulnerability limitations.
Although some work has begun to explore LLMs' repair performance, their effectiveness is unsatisfactory.
To address these limitations, we conducted a large-scale empirical study to investigate the performance of automated vulnerability repair approaches and state-of-the-art LLMs across seven programming languages.
Results show GPT-4o, instruction-tuned with few-shot prompting, performs competitively against the leading approach, VulMaster.
Additionally, the LLM-based approach shows superior performance in repairing unique vulnerabilities and is more likely to repair the most dangerous vulnerabilities.
Instruction-tuned GPT-4o demonstrates strong generalization on vulnerabilities in previously unseen language, outperforming existing approaches.
Analysis shows Go consistently achieves the highest effectiveness across all model types, while C/C++ performs the worst.
Based on findings, we discuss the promise of LLM on multilingual vulnerability repair and the reasons behind LLM's failed cases.
This work takes the first look at repair approaches and LLMs across multiple languages, highlighting the promising future of adopting LLMs for multilingual vulnerability repair.
\end{abstract}



\maketitle

\section{Introduction}
\label{sec:introduction}

Software vulnerabilities are security flaws, glitches, or weaknesses within software code that can be exploited by attackers to compromise the overall security of a system~\cite{nist_csrc_vulnerability}.
For example, the \textit{Log4Shell} vulnerability (CVE-2021-44228), widely regarded as the most critical vulnerability of the last decade, enables attackers to execute malicious code on any affected system, causing severe financial losses.
According to the Common Vulnerabilities and Exposures~\cite{cve_metrics}, the number of software vulnerabilities reported in 2023 reached a record high of 28,961, representing a significant 15.57\% increase compared to 2022.
However, addressing software vulnerabilities frequently necessitates specialized expertise, and the manual resolution of these vulnerabilities is a labor-intensive process~\cite{ji2018coming, forsgren20212020}.
In particular, the time required to fix a vulnerability typically exceeds at least 45 days on average, as indicated by the recent report ~\cite{Edgescan2024Stats}.
Consequently, there is a pressing need for automated approaches to vulnerability repair.

A variety of Deep Learning (DL)-based approaches in recent years have emerged to automate the vulnerability repair process.
These methods learn embeddings of vulnerable programs and generate repair patches accordingly.
Specifically, the attention-based transformer architecture and pre-trained language models (PLMs) have been widely adopted for automated vulnerability repair, demonstrating effective performance~\cite{berabi2021tfix, fu2022vulrepair, zhou2024out, fu2024vision}. 
Their effectiveness primarily stems from the self-attention mechanism, which learns global dependencies among code embeddings, and the pre-trained knowledge that enhances downstream tasks.
For instance, \citet{fu2022vulrepair} introduced VulRepair, a T5-based approach that uses pre-training and Byte-Pair Encoding. 
\citet{chen2022neural} created VRepair, which addresses the scarcity of vulnerability-fixing datasets through transfer learning. 
Most recently, \citet{zhou2024out} presented VulMaster, which builds on CodeT5 and incorporates syntax trees and CWE knowledge. 
VulMaster improves performance by integrating the Fusion-in-Decoder (FiD) framework with multi-task learning, proving to be the most effective approach.

\begin{table}[b]
    \caption{Programming languages targeted by existing repair approaches}
    \label{tab:1_dataset}
    \centering
    \tabcolsep=1.8mm
    \small
   \begin{tabular}{ccc}
       \toprule
            Approach & Studied Dataset & Languages \\
        \midrule
            TFix~\cite{berabi2021tfix} & TFix dataset~\cite{berabi2021tfix} & JavaScript \\
            VRepair~\cite{chen2022neural} & CVEfixes~\cite{bhandari2021cvefixes} and Big-Vul~\cite{fan2020ac} & C/C++ \\
            VulRepair~\cite{fu2022vulrepair} & CVEfixes and Big-Vul & C/C++ \\
            VQM~\cite{fu2022vulrepair} & CVEfixes and Big-Vul & C/C++ \\
            VulMaster~\cite{zhou2024out} & CVEfixes and Big-Vul & C/C++ \\
        \bottomrule
    \end{tabular}
\end{table}

While DL-based vulnerability repair approaches show promise to some extent, their effectiveness is confined to specific programming languages.
Table~\ref{tab:1_dataset} presents the datasets employed in popular DL-based approaches, with CVEfixes~\cite{bhandari2021cvefixes} and Big-Vul~\cite{fan2020ac} being the most frequently utilized.
Notably, both of them focus exclusively on C and C++ languages. 
\textit{A significant limitation is the lack of knowledge about the performance of automated vulnerability repair approaches in a multilingual context} (\textbf{Limitation 1}).
This is particularly important for widely-used languages like Python and Java, found to contain numerous vulnerable codes in open-source projects~\cite{alfadel2023empirical, hu2024empirical}.
\citet{li2022vulnerability} also suggested the importance of assessing and defending against multilingual vulnerabilities.
This indicates that modern software development increasingly utilizes multiple programming languages, leading to diverse codebases with unique security challenges. 
Such complexity makes it difficult for developers to consistently detect and repair vulnerabilities in multilingual programming language environments manually. 
Therefore, there is an urgent need to expand research on automated vulnerability repairs beyond single programming languages such as C and C++.

Large Language Models (LLMs), trained on the large-scale text and code corpus with billions of parameters, have been extensively explored in the software engineering domain and have shown remarkable performance across various code-related tasks~\cite{hou2023large, wang2024software}.
Moreover, current AVR techniques often rely on detailed information, such as the exact CWE type and precise vulnerability locations, while being constrained by a limited number of context tokens for patch generation.
These limitations hinder their applicability in real-world scenarios, where such detailed information is rarely available and a broader contextual understanding is required.
In contrast, LLMs could take only the vulnerable code as input and generate the repaired version without requiring additional information or facing similar output constraints.
Meanwhile, some researchers have already begun exploring the adoption of LLMs into automated vulnerability repair.
For instance, \citet{pearce2023examining} conducted a large-scale study of five commercially available LLMs, examining their capability for zero-shot vulnerability repair. 
Likewise, \citet{10479409} explored the feasibility of two ChatGPT models in prompt-based vulnerability repair.
\textit{Despite these attempts, the role of LLMs in automated vulnerability repair remains largely unexplored} (\textbf{Limitation 2}). 
Specifically, existing LLM-based approaches fall short in two key areas: they are inadequately evaluated in complex development environments (i.e., multilingual code) and lack diverse strategies for invoking LLMs, which leads to unsatisfactory repair performance.

To address the above limitations, we conduct an empirical study to comprehensively explore the effectiveness of the existing automated vulnerability approaches and the potential of leveraging LLMs for repairing vulnerabilities in the multilingual context.
Given LLMs' strong semantic understanding and language-agnostic capabilities, we assume that LLMs paired with effective learning strategies could achieve promising results.
We evaluate the aforementioned approaches using REEF, the latest multi-language vulnerability dataset, which comprises 4,466 CVEs with 30,987 patches across seven popular programming languages (C, C\#, C++, Go, JavaScript, Java, and Python).
We formulate the following four research questions to guide our study:
\\
\noindent
\textbf{RQ1: What is the performance of existing learning-based repair techniques in multilingual vulnerability?}
We first aim to investigate the effectiveness of the existing repair techniques, encompassing both state-of-the-art DL-based approaches and widely used pre-trained language models. 
These techniques are typically devised for a specific programming language in the prior work, hence answering this RQ would provide a better understanding of their knowledge transfer capabilities in vulnerability repair across different languages.
\\ 
\noindent
\textbf{RQ2: What is the performance of state-of-the-art LLMs in repairing multilingual vulnerability?}
This RQ examines the effectiveness of both open-source and closed-source LLMs in repairing multilingual vulnerabilities. 
Different strategies such as code embedding and prompting, utilized by LLMs may influence repair performance. 
Thus, we further investigate the extent of their impact. 
Building on this and drawing inspiration from~\citet{Mueller2024MultiTaskTM}, we design an instruction-tuning strategy as a multi-task learning paradigm to enhance LLMs' ability to capture vulnerability semantics across different programming languages.
Answering this RQ could offer insights into the optimal selection of strategies for enhancing LLM performance on multilingual vulnerability repair.
\\
\noindent
\textbf{RQ3: What are the strengths and weaknesses of the studied automated vulnerability techniques?}
Certain repair techniques and LLM strategies may be prone to repairing unique vulnerabilities based on specific programming languages and vulnerability types.
Therefore, RQ3 seeks to gain an understanding of the characteristics of various repair categories and LLM strategies by analyzing the orthogonality between them, revealing the strengths and weaknesses of different techniques and strategies.
\\
\noindent
\textbf{RQ4: What is the generalization capability in repairing previously unseen vulnerabilities?}
This RQ investigates the effectiveness of the best-performing LLM and AVR techniques in repairing previously unseen vulnerabilities.
Although current methods demonstrate strong performance on in-domain data, their ability to handle out-of-distribution vulnerabilities remains unclear.
Therefore, RQ4 aims to evaluate the generalization capability of these techniques and provide insights into the optimal selection of approaches for addressing unseen vulnerabilities.

Our key empirical findings are: 
(1) Among state-of-the-art AVR and PLM techniques, VulMaster achieves the best performance across the studied evaluation metrics, with an Exact Match score of 28.94\%; 
(2) Among five advanced LLMs with four compositional strategies, the instruction-tuned ChatGPT-4o with a few-shot prompting strategy significantly outperforms others in Exact Match, BLEU, and ROUGE metrics (with an Exact Match score of 28.71\%), which is competitive with the VulMaster; 
(3) Regardless of model type, performance remains relatively consistent across the seven studied programming languages, with Go yielding the best results and C/C++ the poorest;
(4) The orthogonality analysis reveals that the best-performing LLM-based model demonstrates superiority in unique correct repairs and is more likely to repair the most dangerous vulnerabilities compared to VulMaster.
(5) The results on TypeScript vulnerabilities indicate that GPT-4o, when enhanced with instruction tuning and few-shot prompting strategies, exhibits superior generalization to previously unseen programming language vulnerabilities compared to VulMaster.
In addition, our manual analysis shows that when instruction-tuned ChatGPT-4 fails to repair vulnerabilities using few-shot prompting, the main challenge lies in error localization.
\\
\noindent
\textbf{Contributions.}
To sum up, the contributions of this study are:
\begin{enumerate}
    \item The first empirical study to systematically investigate the effectiveness of existing AVR techniques and various LLMs on multilingual vulnerability repair.
    \item The empirical results confirm the promising role of LLMs in multilingual vulnerability repair, particularly when instruction-tuning LLMs with few-shot prompting strategies.
    \item We provide valuable insights into the capabilities and limitations of LLMs for multilingual vulnerability repair, served as essential guidance for future research aimed at enhancing LLM-based vulnerability repair.
    \item We open source all data, code, and analysis details involved in our study\cite{AVR2025}.
\end{enumerate}

\section{Background and Related Work}
\label{sec:background}

\subsection{Software Vulnerability and Datasets}
\label{subsec:soft_vul}
A software vulnerability is a flaw or weakness in a software system that attackers can exploit.
Its impact can be severe, as unpatched vulnerabilities in widely used software may result in catastrophic consequences, including significant economic losses~\cite{bilge12empirical}.
To classify different types and instances of vulnerability, Common Weakness Enumeration (CWE) and Common Vulnerability Exposure (CVE) are used to refer to the types of software weakness that can lead to vulnerabilities and a specific instance of a vulnerability in a software system~\cite{cwe, cve}.
Software vulnerabilities are frequently found in open-source projects.
\citet{alfadel2023empirical} provided insights into common security issues within the Python ecosystem 
by analyzing 550 vulnerability reports affecting 252 Python packages.
~\citet{hu2024empirical} examined the prevalence and remediation delays of vulnerabilities in Go modules, revealing that 66.10\% of modules are affected and identifying two types of lags that hinder timely vulnerability fixes. 
~\citet{meng2018secure} investigated developers' challenges in implementing secure Java coding practices, highlighting vulnerabilities like insecure configurations and misused security libraries, while offering recommendations for improving Java application security. 
~\citet{zhang2021study} examined 646,716 C/C++ code snippets from Stack Overflow, finding that 2\% contained security weaknesses (primarily improper memory operations and null pointer dereferences) and noted that while code revisions often reduced these issues, many weaknesses persisted.

Researchers have developed several datasets to study vulnerability repair.
The VulnLoc ~\cite{shen2021localizing} dataset consists of 43 vulnerable programs in 10 projects that span 6 CWEs.
The Vul4J ~\cite{bui2022vul4j} dataset encompasses 79 reproducible vulnerabilities drawn from 51 open-source Java projects, spanning 25 different CWE types.
\citet{reis2021ground} developed a dataset of 5,942 security patches from the complete CVE details database, spanning 1,339 projects and 146 vulnerability types across 20 languages.
SATE IV~\cite{okun2013report} is a dataset initially designed to assess static analysis tools for detecting security-relevant defects which consists of 117,738 synthetic C/C++ functions categorized under 116 CWE types.
VulinOSS~\cite{gkortzis2018vulinoss} is a dataset derived from open-source projects, assembled using vulnerability reports from the National Vulnerability Database (NVD), encompassing 17,738 vulnerabilities across 153 projects.
Big-Vul~\cite{fan2020ac} is a dataset of C/C++ code vulnerabilities sourced from 348 open-source GitHub projects which comprises 3,754 vulnerabilities and represents 91 distinct CWE types.
This dataset is designed for multiple purposes, including vulnerability detection, vulnerability repair, and vulnerability analysis.
CVEfixes~\cite{bhandari2021cvefixes} is a vulnerability dataset based on CVE records up to June 9, 2021 from the NVD, containing 5,365 CVEs from 1,754 projects.
Recently, \citet{wang2023reef} proposed REEF, an automated framework for collecting high-quality vulnerabilities and their fixes across various languages, platforms, and granularity.
The dataset details are shown in Section~\ref{subsec:dataset}.

\subsection{Automated Vulnerability Repair}
Researchers have proposed leveraging various Neural Machine Translation (NMT) approaches for Automated Vulnerability Repair (AVR).
\citet{NEURIPS2018_68abef8e} introduced using generative adversarial networks (GAN) for AVR task.
They utilized a traditional NMT model as the generator to generate the examples to confuse the discriminator whose task is to discern the NMT-generated repairs from the real repairs.
\citet{chen2022neural} proposed VRepair which leverages a word-level tokenizer and a vanilla Transformer model.
VRepair starts by pre-training the vanilla Transformer model on a bug-fixing corpus and subsequently uses it to address C/C++ vulnerabilities with fine-tuning.
Similar to VRepair, \citet{chi2022seqtrans} introduced SeqTrans, a Transformer-based NMT model with copy mechanisms designed to repair Java vulnerabilities.
\citet{fu2022vulrepair} introduced VulRepair, a system that utilizes a Byte Pair Encoding (BPE) tokenizer and CodeT5, which is pre-trained on a substantial corpus of code to address C/C++ vulnerabilities.
\citet{wu2023effective} conducted an evaluation involving four program repair models and nine large language models on the Vul4J dataset.
\citet{zhou2024out} proposed VulMaster. This Transformer-based NMT model addresses the input length constraints inherent in traditional Transformer-based pre-trained models by employing the Fusion-in-Decoder (FiD) framework~\cite{izacard2021leveraging}.
By eliminating the length limit, VulMaster can integrate vulnerable code structures and expert knowledge to further enhance its repair capabilities on repairing C/C++ vulnerabilities.
Inspired by Vision Transformer (VIT)-based objection detection approaches~\cite{carion2020end, zhu2020deformable} in the computer vision domain, \citet{fu2024vision} introduced a T5-based approach named VQM designed to enhance awareness and attention to vulnerable code areas within a function to produce better repairs for C/C++ vulnerabilities.
According to the literature, while various AVR techniques exist, they primarily focus on C/C++ as their target language.

With the rapid advancement of LLMs, research has increasingly focused on evaluating their effectiveness in vulnerability repair.
\citet{pearce2023examining} examined the power of zero-shot prompting on vulnerability repair, while
\citet{10479409} instructed ChatGPT for repairing vulnerabilities over extensive real-world C/C++ datasets.
\citet{kulsum2024case} proposed an LLM-based vulnerability repair technique based on chain-of-thought prompt and patch validation feedback
on C vulnerabilities of CVEfixes dataset. 
\citet{zhang2024evaluating}  evaluated the capability of advanced LLMs in fixing memory corruption vulnerabilities in real-world C/C++ code. 
\citet{fakih2025llm4cveenablingiterativeautomated} proposed an LLM-based iterative pipeline designed to effectively and robustly repair vulnerable functions in C code.
\citet{zhou2024large} conducted a systematic literature review of existing LLM-based approaches for vulnerability repair.
Despite these efforts, the effectiveness of LLMs in automated vulnerability repair remains largely unexplored due to the following limitations: (i) the performance is unsatisfactory yet, as demonstrated in \citet{10479409}; (ii) the diverse learning strategies have not been utilized to unlock the LLMs' potential; and (iii) evaluated datasets are relatively small and do not cover a wide range of programming languages. 
Our study addresses these limitations by comprehensively evaluating how popular and advanced LLMs perform using different strategies for automated vulnerability repair. 
To better understand how far we are, we compare their performance against state-of-the-art AVR techniques and PLMs, using fine-grained analysis to investigate their characteristic differences.

\section{Study Design}
\label{sec:evaluation_design}

\begin{figure}[t!]
    \centering
    \includegraphics[width=\linewidth]{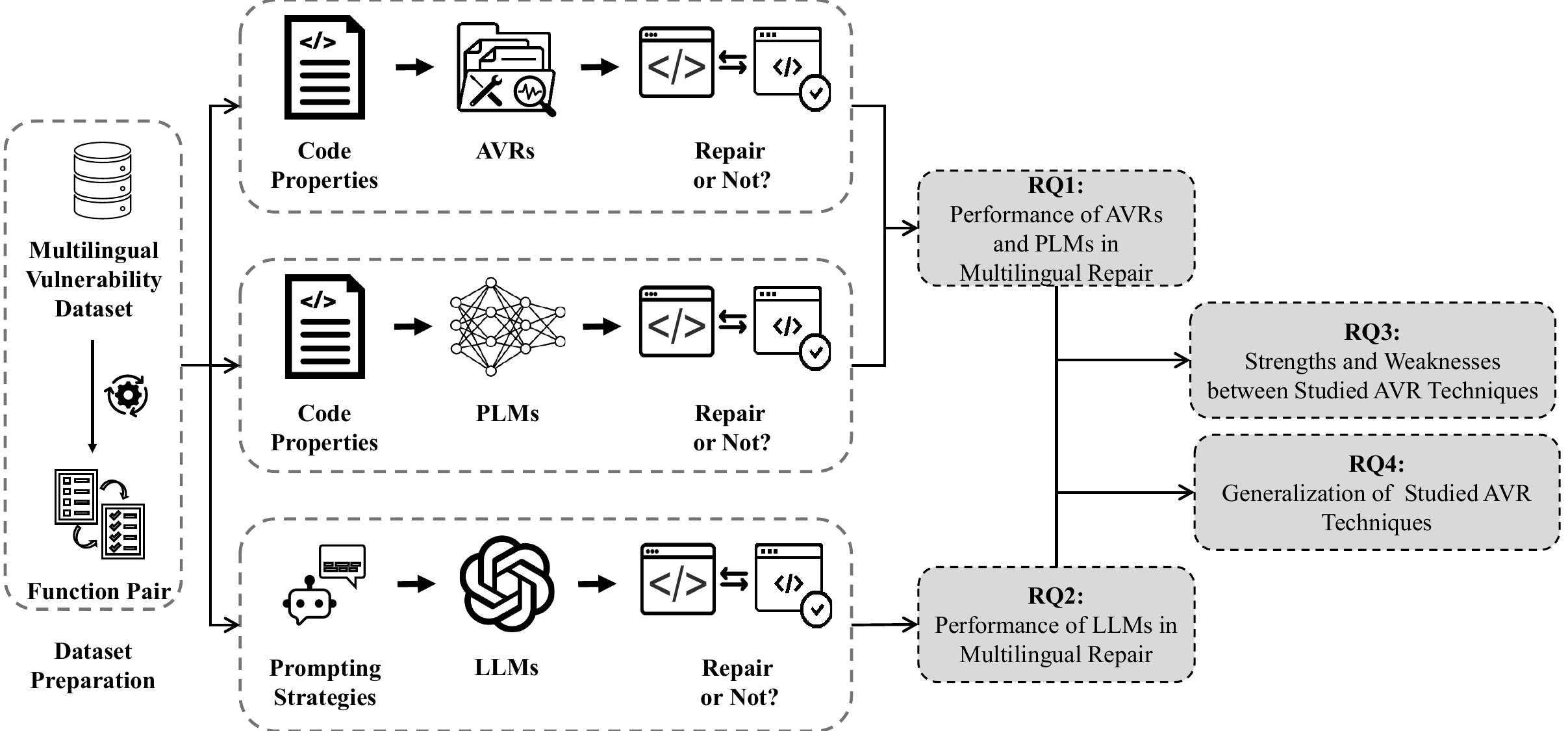}
    \caption{Overview of study design}
    \label{fig:overview}
\end{figure}

Figure ~\ref{fig:overview} illustrates the overview of our study design.
Initially, based on the latest multi-language vulnerability dataset REEF ~\cite{wang2023reef}, we construct the training and test datasets comprising code pairs of the vulnerability function and its repair.
We first investigate the effectiveness of the existing learning-based AVR techniques and widely used PLMs on automated vulnerability repair in the multilingual context (RQ1).
We also examine how different LLM strategies affect the performance of automated vulnerability repair (RQ2).
Finally, we gain insights into the strengths and limitations of the studied AVR techniques by analyzing their orthogonality (RQ3) and further investigate their generalization capabilities (RQ4).

\subsection{Dataset Preparation}
\label{subsec:dataset}

\textbf{Studied dataset}. 
To evaluate the performance of AVR techniques, we select the multi-language vulnerability dataset REEF ~\cite{wang2023reef}, containing 4,466 CVEs with 30,987 patches (including 236 CWE) across seven programming languages with detailed related information including vulnerability information (e.g., the identifier of CVE and the CVSS scores reflecting the severity) and project information (e.g., the commit message and the URL of the source file changed in the commit).
More specifically, REEF gathers real-world vulnerabilities from the NVD database and CVE list maintained by Mend ~\cite{whitesource2022mend} which is an open-source vulnerabilities database and collects CVEs from 2016 to 2023. 
In this study, we focus on all seven programming languages: C, C++, C\#, Go, Java, JavaScript, and Python.
At the function level, the dataset comprises 6,957 C, 2,244 C++, 1,529 C\#, 3,187 Go, 6,207 Java, 5,066 JavaScript, and 5,797 Python functions.

\textbf{Data pre-processing}. 
Since the REEF dataset was not originally collected for automated vulnerability repair tasks, we needed to first retrieve the vulnerable functions.
The commit data comes in a raw code format with patches that cannot be directly used for extracting vulnerable functions. To obtain the commit data containing both pre-change and post-change files, we applied patches to the raw code using the Linux patch command.~\footnote{\url{https://www.man7.org/linux/man-pages/man1/patch.1.html}}
After obtaining the required commit data, we removed all code comments to reduce potential bias and collected the function definition code,  utilizing the static analysis tool Tree-sitter~\cite{brunsfeld2024tree}.
Tree-sitter is a parser generator tool and incremental parsing library that can be used to parse any programming language.
To extract the vulnerable function definition and corresponding repair code, we iterated the function definition in the pre-change file in order to match the corresponding post-change function definition.
Specifically, the function name is used as a key to match the same function definition in the post-change file.
If multiple function definitions are matched in the post-change file, the edit distance would be calculated between the pre-change function definition and each post-change function definition, and the function definition pair with the minimal edit distance is selected as the vulnerable function and its repair.
In other words, we define the pre-change function and the post-change function with different source codes as a function pair. 
After this step, we obtained a total of 11,808 function pairs. 

To minimize bias from potential data leakage, we removed function pairs that appeared in both the Big-Vul and CVEfixes datasets, resulting in 1,159 duplicate pairs.
In the end, we were able to collect 10,649 vulnerable function pairs for seven languages, as shown in Table~\ref{tab:dataset}. 
The reduction in vulnerable functions compared to the original dataset could be attributed to three main reasons: (i) the Tree-sitter cannot parse certain cases (such as special C macros), (ii) some functions have multiple patches, and (iii) some functions are completely deleted or added.
Moreover, we performed a sanity check to ensure the robustness of the Tree-sitter. 
To do so, the first author randomly selected 50 samples to examine both the consistency of function definitions between pre-change and post-change versions and their alignment with the provided patch.
The manual results suggest that the Tree-sitter is reliable and produces no false positives.

\begin{table}[]
    \caption{Statistic summary of the experimental dataset}
    \label{tab:dataset}
    \centering
    \tabcolsep=2.2mm
    \small
\begin{tabular}{ccccc}
\hline
Languages & Training & Validation & Test & Total \\ \hline
C         & 1,063     & 151        & 305  & 1,519  \\
C++       & 959      & 137        & 275  & 1,371  \\
C\#       & 159      & 22         & 47   & 228   \\
Go        & 1,171     & 167        & 334  & 1,672  \\
Java      & 1,193     & 167        & 343  & 1,703  \\
JavaScript        & 1,668     & 239        & 484  & 2,391  \\
Python    & 1,235     & 176        & 354  & 1,765  \\ \hline
Total     & 7,448     & 1,059       & 2,142 & 10,649 \\ \hline
\end{tabular}
\end{table}

\textbf{Construction of training and test datasets}.
Following existing work~\cite{fu2024vision, zhou2024out}, we adopted a widely-used random sampling strategy to divide the REEF dataset into training, validation, and test sets with a ratio of 7:1:2.
To ensure that data across different languages maintain consistent proportions, we first allocate the data for each language based on the predetermined ratio. 
Subsequently, we collect the corresponding portions to assemble the final dataset.
As shown in Table~\ref{tab:dataset}, our experimental dataset contains 7,448, 1,059, and 2,142 function pairs for training, validation, and testing over seven languages.

\begin{figure}[t!]
    \centering
    \includegraphics[width=.8\linewidth]{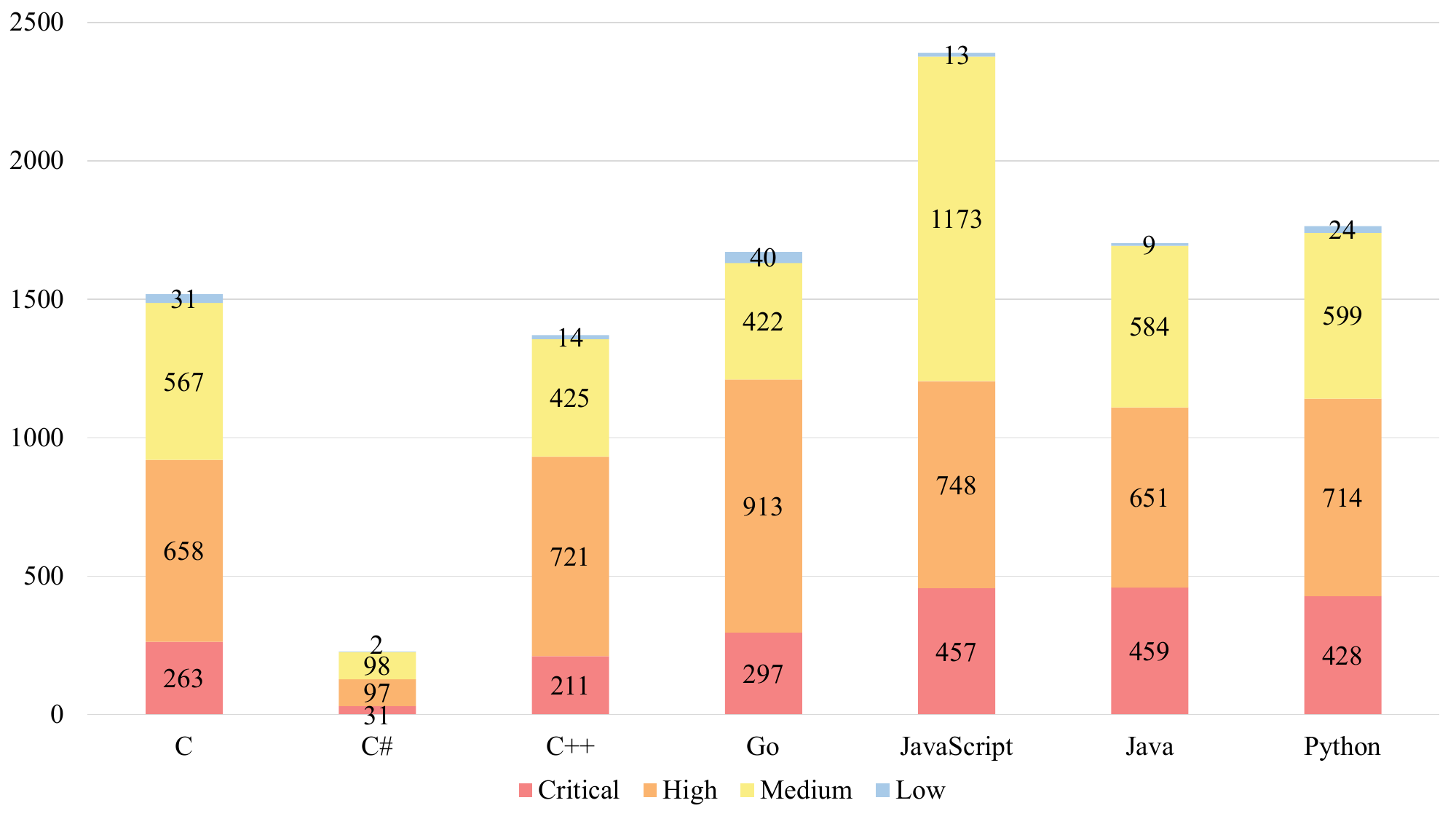}
    \caption{The severity distribution across seven programming languages on the experimental dataset}
    \label{fig:3_severity}
    
\end{figure}

\textbf{Distribution of vulnerability severity.} To analyze the dataset quality, we examined the severity levels of all 10,649 vulnerabilities, as shown in Table~\ref{tab:dataset}.
We relied on the Common Vulnerability Scoring System (CVSS), a method used to supply a qualitative measure of severity.
Specifically, we used the metric CVSS v4.0 Ratings\footnote{\url{https://nvd.nist.gov/vuln-metrics/cvss}} to classify the vulnerabilities into four levels: low, medium, high, and critical.
Figure ~\ref{fig:3_severity} depicts the related severity distribution across the studied seven programming languages.
Analysis shows that only 1.2\% of vulnerabilities are rated as low severity, while a significant 61.3\% are classified as high or critical severity. 
These results indicate that the real-world vulnerabilities provided by REEF are of high quality and more likely to target severe vulnerabilities.

\subsection{Existing Automated Vulnerability Repair Approaches}
\label{subsec:models}
In our study, we investigated the performance of two types of existing approaches for AVR: learning-based automatic vulnerability repair approaches and pre-trained language models.
For learning-based approaches, we included five state-of-the-art approaches.

\begin{itemize}
    \item \textbf{TFix}~\cite{berabi2021tfix} is a T5-based approach that utilizes a large language model pre-trained on a natural language corpus ~\cite{raffel2020exploring}.
    
    \item \textbf{VRepair}~\cite{chen2022neural} is a vanilla transformer architecture with word-level tokenization for input, pre-trained on a bug-fixing~\cite{chen2022neural} dataset and fine-tuned for AVR. 
    
    \item \textbf{VulRepair}~\cite{fu2022vulrepair} is a T5-based approach that incorporates the Byte Pair Encoding (BPE) method~\cite{sennrich2015neural} for subword-level tokenization, using a large language model pre-trained on a source code corpus~\cite{wang2021codet5}.  

    \item \textbf{VulMaster}~\cite{zhou2024out} is a transformer-based neural network model that  integrates source code with code syntax trees and specific CWE knowledge in its inputs, employing Fusion-in-Decoder (FiD) framework~\cite{izacard2021leveraging} and multi-task learning framework~\cite{wang2022test}.

    \item \textbf{VQM}~\cite{fu2024vision} is a T5-based approach that utilizes vision transformer methodologies to enhance the encoder and decoder focus on vulnerable areas of code during the repair process.
    
\end{itemize}

For pre-trained language models, we selected seven models across three architectures (encoder-only, decoder-only, and encoder-decoder) that are widely used in the literature ~\cite{fu2022vulrepair, zhou2024out, fu2024vision}, including:
\begin{itemize}
    \item \textbf{CodeBERT}~\cite{feng2020codebert} is a widely recognized pre-trained model that leverages a multilayer Transformer architecture to effectively learn from bimodal data, comprising both source code and natural languages.
    
    \item \textbf{GraphCodeBERT}~\cite{guo2020graphcodebert} is a pre-trained model that utilizes a transformer-based architecture to enhance code representation by incorporating semantic-level information from code.

    \item \textbf{CodeT5}~\cite{wang2021codet5}  is a unified encoder-decoder model that extends the T5 architecture by incorporating token type information in code and utilizing denoising sequence-to-sequence pre-training to improve model performance. 

    \item \textbf{CodeReviewer}~\cite{li2022automating} is a pre-trained encoder-decoder model specifically designed for analyzing code diffs and reviews, featuring four tailor-made pre-training tasks to enhance its effectiveness in the code review scenario.

    \item \textbf{PolyCoder}~\cite{xu2022systematic} is a pre-trained GPT-2-based model~\cite{radford2019language} specifically designed for multi-lingual code modeling, developed to address the gap in large open-source models exclusively trained on code.

    \item \textbf{CodeGen}~\cite{nijkamp2022codegen} is a pre-trained decoder-only model which specifically engineered for program synthesis using natural and programming language data.

    \item \textbf{GPT2-CSRC}~\cite{pearce2023examining} is a pre-trained decoder-only model based on GPT-2, equipped with a BPE tokenizer and specifically developed for C/C++ code analysis, having been trained on a substantial dataset of approximately 17 GB of C/C++ code.

\end{itemize}

\subsection{Experimented Large Language Models}
We studied five advanced LLMs (i.e., three open-source LLMs and two closed-source LLMs), which have demonstrated excellent performance in the various code-related downstream tasks~\cite{wang2024software}. The details of these LLMs are as follows: 

\begin{itemize}
       
    \item 
    \textbf{DeepSeek-Coder}~\cite{guo2024deepseek} trained on 2 trillion tokens across 87 programming languages, utilizing a 16K context window and fill-in-the-blank tasks to enhance code generation and infilling capabilities, achieving state-of-the-art performance among open-source models and surpassing closed-source counterparts like Code Llama and GPT-3.5 in some tasks.

    \item \textbf{Code Llama}~\cite{roziere2023code} is a decoder-only model which is one of the most popular LLMs for code generation and infilling derived from Llama 2 models.
    It undergoes additional fine-tuning on 500B tokens derived from an extensively code-rich dataset.

    \item \textbf{Llama 3}~\cite{dubey2024llama} is a series of large multilingual models based on the Transformer architecture, designed to improve performance across various language understanding tasks. 
    By optimizing data quality, training scale, and model architecture, Llama 3 demonstrates significant potential in the field of natural language processing. 
    
    \item \textbf{ChatGPT}~\cite{chatgpt2022} is a groundbreaking  LLM capable of transforming various fields through its advanced natural language processing capabilities.
    It is trained on a vast corpus of natural language texts and code snippets, employing reinforcement learning to enhance its ability to adhere to human directives.
    Specifically, we studied two LLMs, i.e., GPT-3.5-Turbo~\cite{chatgpt2022} and GPT-4o~\cite{openai2024gpt4o}.

\end{itemize}

\subsection{Strategies for Large Language Models}
\label{subsec:strategy}
\begin{figure}[t]
    \centering
    \begin{subfigure}{0.8\linewidth}
        \centering
        \includegraphics[width=\linewidth]{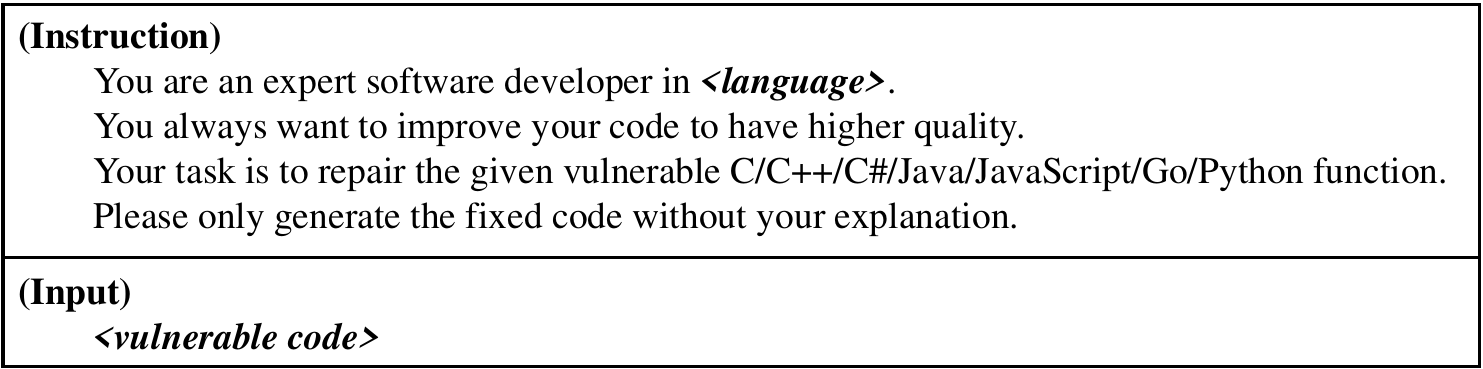}
        \caption{\small A prompt template for zero-shot prompting}
        \label{fig:zsp}
    \end{subfigure}
    
    \vskip 0.5em

    \begin{subfigure}{0.8\linewidth}
        \centering
        \includegraphics[width=\linewidth]{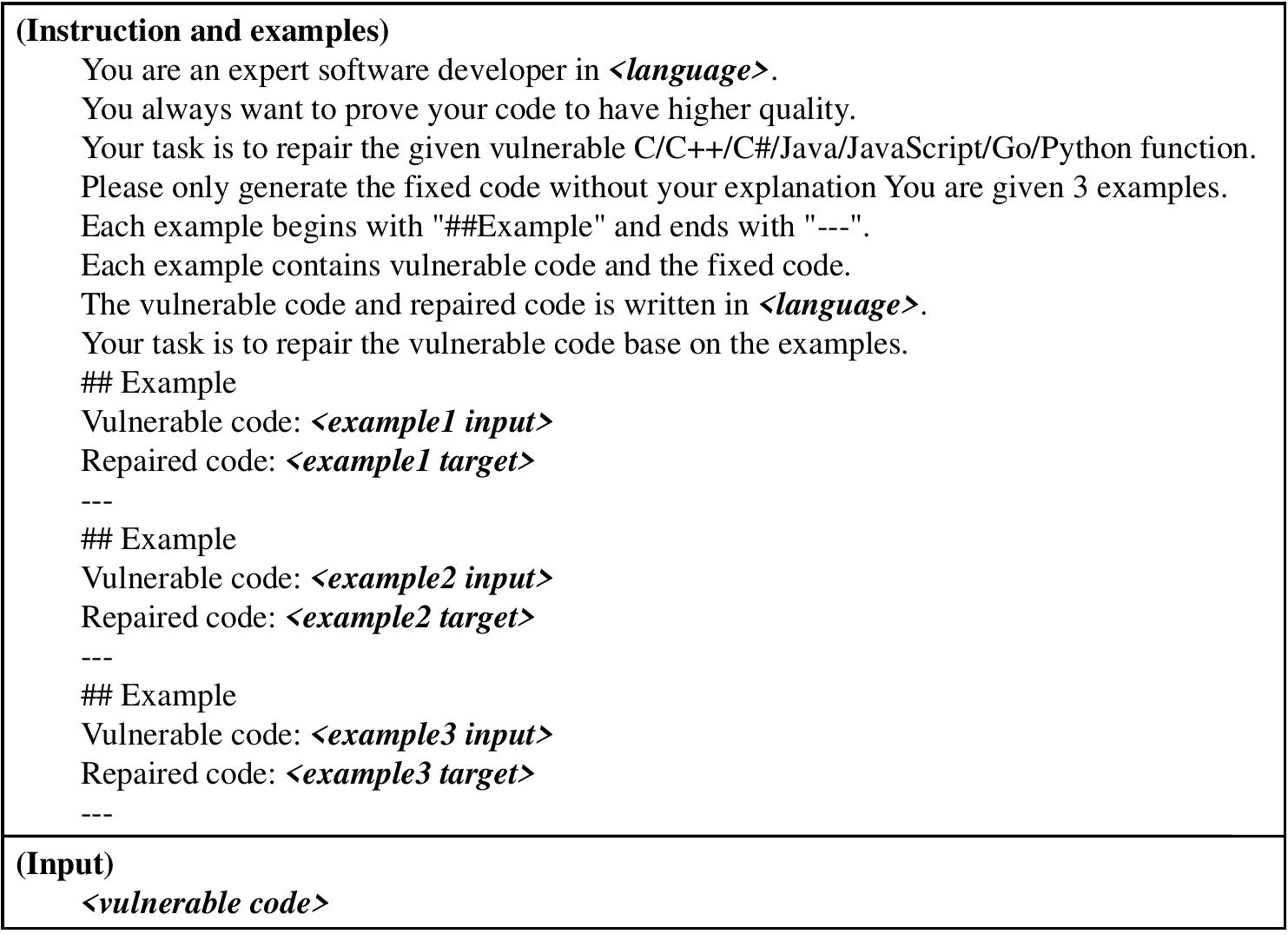}
        \caption{\small A prompt template for few-shot prompting}
        \label{fig:fsp}
    \end{subfigure}
    
    \caption{Zero-shot prompt and few-shot prompt for multilingual AVR}
    \label{fig:prompt}
\end{figure}

We further investigated the impact of LLMs' learning strategies in automated vulnerability repair.
Existing studies~\cite{tian2024large} have demonstrated that fine-tuning strategies can lead to significant enhancement in LLM performance effectively by adapting general LLMs to specific downstream tasks.
Meanwhile, prompting strategies have also been proposed to achieve the same objective in a plug-and-play manner~\cite{kojima2022large}.
Thus, we devised three LLM strategies as follows:
\begin{itemize}[leftmargin=10pt]
   
    \item \textbf{Zero-shot prompting strategy}: we devised a prompt with system role and user role following prior works~\cite{zhou2024out, zhang2023pre} without any examples, which directly utilized a structured instruction and a vulnerable function to prompt LLMs for vulnerability repair. Figure~\ref{fig:zsp} shows the prompt template for zero-shot prompting. In our prompt design, we first assign a specific role to the LLM (e.g., ``You are an expert software developer in <language>"), then define the vulnerability repair task, and finally request the LLM to generate the repaired code.

    \item \textbf{Few-shot prompting strategy}: it enables LLMs to learn the relationship between the vulnerable function and the repaired function based on selected function pair examples. 
    That is, it concatenates these demonstration examples with a zero-shot prompt to form a new few-shot prompt, which is then fed to LLMs for vulnerability repair.
    Figure~\ref{fig:fsp} shows the prompt template for few-shot prompting. 
    Following prior study ~\cite{PORNPRASIT2024107523}, for a given vulnerable function, three demonstration examples from the REEF training dataset are selected by BM25~\cite{robertson2009probabilistic} to prompt LLMs for vulnerability repair. 
    BM25 is selected as the sample selection approach since prior work~\cite{gao2023makes, yuan2023evaluating} shows that BM25 outperforms other sample selection approaches for software engineering tasks.
    We choose three demonstration examples for each function sample based on the findings of \citet{gao2023makes}, which indicate that GPT-3.5 can achieve approximately 90\% of its highest Exact Match score with just three examples, compared to using 16 or more.
    
    \item \textbf{Instruction-tuning strategy}: it enables LLMs to acquire specific knowledge through training on many more instruction-filled function pairs.
    Research suggests that fine-tuning LLMs on diverse multi-task datasets accompanied by natural language descriptions significantly enhances their performance on previously unseen tasks\cite{ouyang2022training}.
    Specifically, we used the same structured instruction that is described in the previous zero-shot prompting strategy to construct an instruction-filled fine-tuning set.
    Then, we fine-tuned the LLMs on the fine-tuning set to repair vulnerable functions by the zero-shot prompt or few-shot prompt.

\end{itemize}


\subsection{Evaluation Metrics}
\label{subsec:metrics}

In accordance with previous studies~\cite{chen2022neural, fu2022vulrepair, zhang2023pre, zhou2024out, fu2024vision}, we used the most widely-adopted \textit{Exact Match (EM)} metric with beam search to evaluate our automated vulnerability repair techniques.
A repair is considered an EM if any beam search output matches the ground-truth label exactly. We evaluated all methods using beam sizes of 1, 3, and 5.

We also utilized another two widely-used metrics for code generation tasks, i.e. BLEU-4 ~\cite{papineni2002bleu} and ROUGE~\cite{lin2004rouge}.
BLEU-4 measures how similar the generated code is to the ground truth at the token level. It focuses on precision by comparing n-grams and applying length penalties.
In contrast, ROUGE evaluates how well the generated code overlaps with the ground truth, with a particular focus on recall.
We adopted ROUGE-1, ROUGE-2, and ROUGE-L as metrics.
These metrics evaluate the quality of generated texts from distinct perspectives and are often used in conjunction to provide a more comprehensive assessment.

\subsection{Implementation and Environment}
\label{subsec:implementation}
We replicated TFix, VRepair, VulRepair, VQM, and VulMaster by following their documented implementation instructions and parameter settings provided in replication packages.
All the open-source pre-trained models (i.e., CodeBERT, GraphCodeBERT, CodeT5, CodeReviewer, PolyCoder, CodeGen and GPT2-CSRC) are downloaded from Huggingface~\cite{wolf2019huggingface} and we employed the parameter settings suggested by the work of \cite{fu2024vision}.

Regarding closed-source LLMs, we invoked GPT-3.5-Turbo and GPT-4o through OpenAI’s APIs~\cite{openai2024}.
Concretely, we used \texttt{gpt-3.5-turbo-0125} as the specific experimental model version for GPT-3.5-Turbo and used \texttt{gpt-4-0613} as the experimental model version for GPT-4o.
Regarding open-source LLMs (i.e., DeepSeek-Coder, Code Llama, and Llama 3), we downloaded the pre-trained model from Huggingface~\cite{wolf2019huggingface}. Specifically, the model sizes are 6.7B, 7B, and 8B for the three studied open-source LLMs, respectively.
To save computational cost and prevent over-fitting, we adopt Low-Rank Adaption (i.e., LoRA~\cite{hu2021lora}) to fine-tune the open-source LLMs.
We configured the LLMs to generate a maximum of 1,024 new tokens, while preserving all other parameters at their default settings.
We conducted all the experiments on an Intel Xeon CPU Gold-6342 machine with 512 GB RAM, Ubuntu 20.04.6, and two A800 GPUs.
More implementation details can be found in our replication package. 
To facilitate future research, we have made publicly available all used datasets and model execution scripts. 

\section{Evaluation Results}
\label{sec:results_and_analysis}

\subsection{RQ1: What is the performance of existing learning-based repair techniques in multilingual vulnerability?}
\label{subsec:RQ1}
\noindent
\textbf{\emph{\underline{Approach.}}}
This research question offers a comparative analysis of the performance between various AVRs and state-of-the-art PLMs in repairing multilingual vulnerabilities.
Specifically, we investigated the effectiveness of five AVRs and seven PLMs.
The model information is presented in Section~\ref{subsec:models}.
We now detail the pre-training process below.

Although PLMs are pre-trained on a large number of natural languages or source code snippets, they have not seen vulnerability repairs with special tokens during their pre-training, resulting in a lack of generating special tokens in vulnerability repairs. 
Prior studies~\cite{chen2022neural, fu2022vulrepair, zhang2023pre, zhou2024out, fu2024vision} have demonstrated that the effectiveness of pre-training on a bug fixing task would facilitate the performance of models on vulnerability repair.
To incorporate code AST information into input, VulMaster is pre-trained on a bug-fixing dataset provided by Chen et al. ~\cite{chen2022neural} which consists of over 500,000 pairs of buggy and fixed functions.
Therefore, to mitigate potential bias, we conducted pre-training with AVRs and PLMs on this bug-fixing dataset, aligning them with each other.

Following existing work~\cite{chen2022neural}, we reuse their code to process vulnerable function pairs and adapt them as inputs for AVR and PLM models.
Particularly, each vulnerable function is marked using the special tokens <StartLoc> and <EndLoc>. The <StartLoc> token indicates the beginning of the vulnerable code lines and <EndLoc> token indicates the end.
To put it another way, <StartLoc> and <EndLoc> provide vulnerability localization to vulnerability repair approaches.
For the repair function, changed codes with three context tokens are extracted as the ground truth which are surrounded with special tokens <ModStart> and <ModEnd>.
In other words, vulnerability repair approaches generate repair code segments that can be mapped to the changed before function rather than the whole repair function.
For each vulnerable function, we add CWE-type information at the beginning. For functions without CWE information, we label them as CWE-000.

To evaluate the effectiveness of the studied AVRs and PLMs, we used multiple metrics: Exact Match (EM) with beam sizes of 1, 3, and 5, BLEU score, and ROUGE-1, ROUGE-2, and ROUGE-L metrics, as detailed in Section~\ref{subsec:metrics}.

\begin{table}[t]
    \caption{Performance of existing learning-based techniques on multilingual vulnerability repair}
    \label{tab:rq1}
    \centering
    \tabcolsep=3.0mm
    \small
    \begin{adjustbox}{max width=1.0\textwidth, center}
    \begin{tabular}{lrrrrrrr}
        \toprule
        \toprule
        \textbf{Technique} & \textbf{EM with beam 1} & \textbf{EM with beam 3} & \textbf{EM with beam 5} & \textbf{BLEU-4} & \textbf{ROUGE-1} & \textbf{ROUGE-2} & \textbf{ROUGE-L} \\ 
        \midrule
        \rowcolor{mygray}\multicolumn{8}{l}{\textbf{Learning-based AVR technique}} \\ 
        \midrule
            TFix      & 10.32\% & 12.79\% & 12.70\% & 24.19\% & 0.5129 & 0.3124 & 0.4966 \\
            VRepair   & 3.97\%  & 4.90\%  & 5.32\%  & 14.09\% & 0.4072 & 0.1833 & 0.3909 \\
            VulRepair & 14.33\% & 18.58\% & 19.28\% & 32.18\% & 0.5900 & 0.3943 & 0.5727 \\
            VQM       & 15.83\% & 19.42\% & 20.03\% & 32.76\% & 0.5856 & 0.4029 & 0.5685 \\
            VulMaster & \textbf{28.94\%} & \textbf{33.47\% } & \textbf{34.59\%} & \textbf{42.99\%} & \textbf{0.6753} & \textbf{0.5149} & \textbf{0.6602}\\ 
        \midrule
        \rowcolor{mygray}\multicolumn{8}{l}{\textbf{Encoder-only PLM technique}} \\ 
        \midrule
            CodeBERT      & 9.24\% & 11.95\% & 13.12\% & 26.14\% & 0.5137 & 0.3014 & 0.4934 \\
            GraphCodeBERT & 8.82\% & 11.20\% & 12.14\% & 23.71\% & 0.5125 & 0.2957 & 0.4939 \\ 
        \midrule
        \rowcolor{mygray}\multicolumn{8}{l}{\textbf{Encoder-decoder PLM technique}} \\ 
        \midrule
            CodeT5       & 14.33\% & 18.58\% & 19.28\% & 32.18\% & 0.5900 & 0.3943 & 0.5727 \\
            CodeReviewer & 13.77\% & 17.60\% & 18.35\% & 31.49\% & 0.5777 & 0.3894 & 0.5620 \\
        \midrule
        \rowcolor{mygray}\multicolumn{8}{l}{\textbf{Decoder-only PLM technique}} \\ 
        \midrule
            PolyCoder & 6.40\% & 7.89\%  & 7.52\%  & 16.86\% & 0.4773 & 0.2435 & 0.4592 \\
            CodeGen   & 9.20\% & 11.34\% & 11.62\% & 20.98\% & 0.5142 & 0.2880 & 0.4968 \\
            GPT2-CSRC & 9.80\% & 10.74\% & 11.11\% & 22.22\% & 0.5289 & 0.3054 & 0.5095 \\
        
        \bottomrule
        \bottomrule
    \end{tabular}
    \end{adjustbox}
\end{table}

\begin{figure*}[]
    \centering
    \includegraphics[width=1.\linewidth]{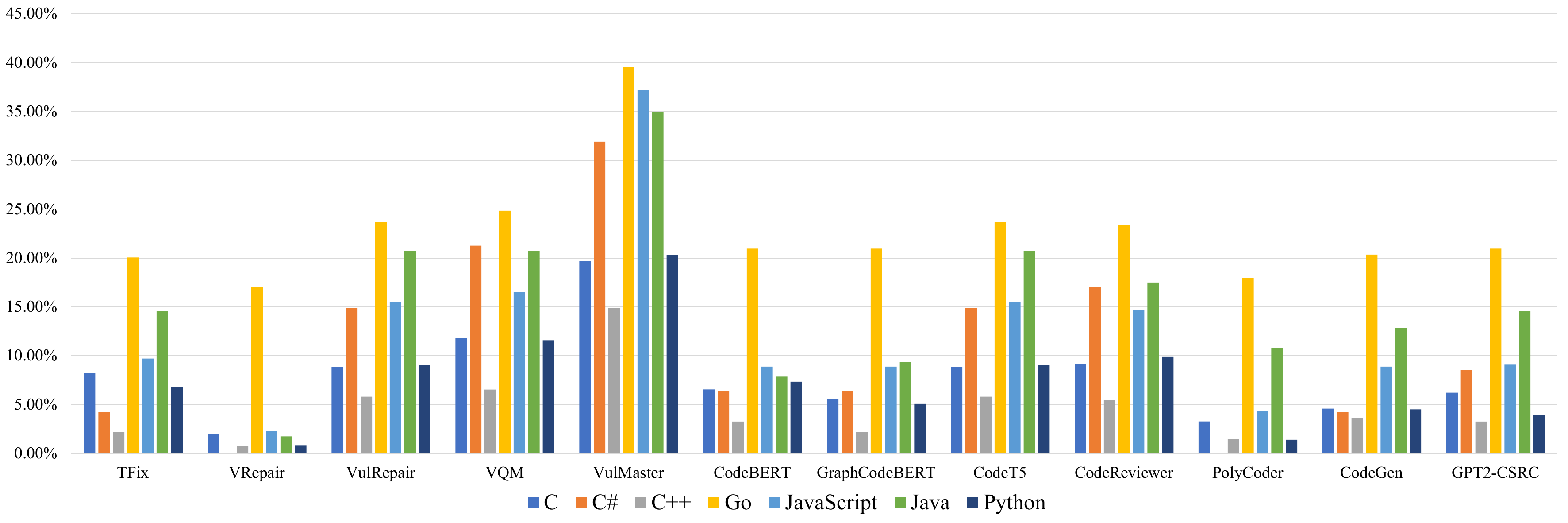}
    \caption{Performance of AVR and PLM techniques across seven programming languages (y-axis: Exact Match score with beam size 1)}
    \label{fig:rq1}
\end{figure*}

\begin{figure*}[]
    \centering
    \includegraphics[width=.7\linewidth]{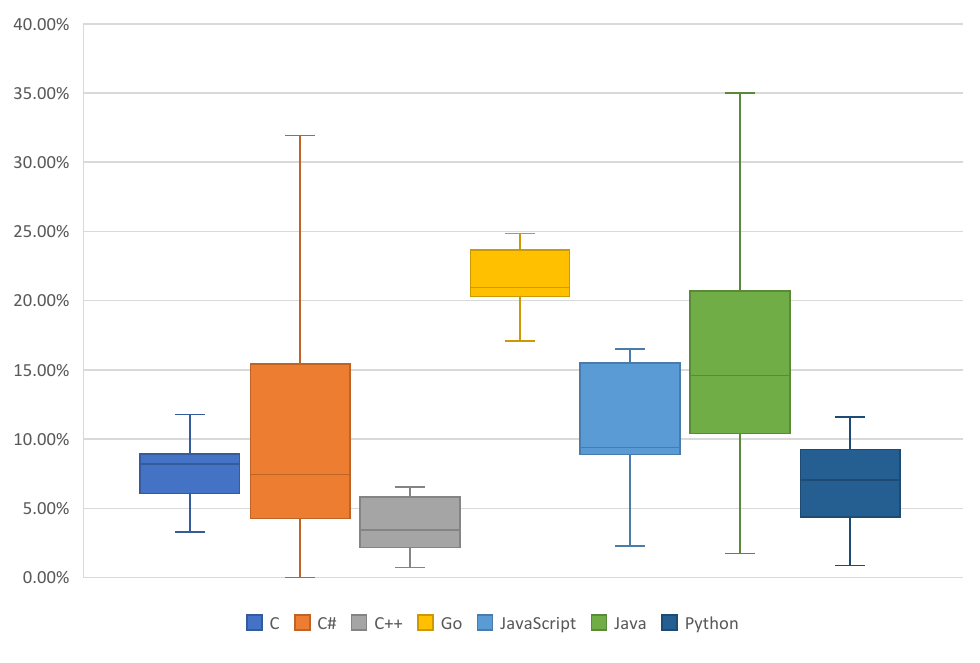}
    \caption{Overall performance of AVR and PLM techniques across seven programming languages (y-axis: Exact Match score with beam size 1)}
    \label{fig:rq1_box}
\end{figure*}

\noindent
\textbf{\emph{\underline{Results.}}}
Table~\ref{tab:rq1} shows the comparison results among AVRs and PLMs in terms of EM, BLEU, and ROUGE scores in the multilingual vulnerability repair context.
Within each metric, the values in bold indicate the technique that exhibits the best performance among all AVRs and PLMs.
Figure~\ref{fig:rq1} and Figure~\ref{fig:rq1_box} present the performance of AVRs and PLMs across seven programming languages in terms of EM with beam size 1.  

VulMaster significantly outperforms other models across all metrics, demonstrating the highest effectiveness in EM with beams 1, 3, and 5, as well as in BLEU, ROUGE-1, ROUGE-2, and ROUGE-L scores.
In particular, VulMaster achieves an impressive 28.94\% on the EM with beam 1 metric, surpassing other methods by a substantial margin ranging from 82.82\% to 628.97\%.
Followed by VulMaster, VQM which is a T5-based approach utilizing vision transformer methodologies achieves an EM of 15.83\% with beam 1.
As for the worst effective technique VRepair in learning-based AVR techniques, it is the only technique that pre-trains a transformer from scratch on the C/C++ only bug-fixing dataset and fine-tuning on the REEF training dataset which comprises seven language programs rather than integrating with powerful PLMs.
Moreover, we observe that VulMaster demonstrates relatively balanced performance across different programming languages as shown in Figure~\ref{fig:rq1}, highlighting its stability in cross-language scenarios.

Encoder-decoder PLM techniques on average outperform both Encoder-only and Decoder-only techniques across all seven evaluation metrics.
Specifically, Encoder-decoder PLM techniques achieve an average improvement of 55.56\% over Encoder-only techniques and 65.99\% over Decoder-only techniques in terms of the EM with beam 1 metric.
Among Encoder-decoder PLM techniques, the best-performing CodeT5 achieves an EM of 14.33\% with beam 1.
The prevalence of this phenomenon may stem from the fact that the automated vulnerability repair task which gives a vulnerability function and outputs corresponding repair closely aligns with code translation which is the pre-training task for the Encoder-decoder PLM.

At the language level, as shown in Figure~\ref{fig:rq1} and Figure~\ref{fig:rq1_box}, the performance of techniques varies across programming languages. While many models show similar performance patterns, Go consistently achieves the best results, while C/C++ shows the relatively weakest performance.
This may be attributed to the relatively recent emergence of the Go programming language and its concise syntax.
VulMaster demonstrates superior performance compared to other AVR and PLM techniques across all seven examined programming languages. It shows heterogeneous efficacy depending on the language, achieving its highest accuracy rates in Go (39.52\%) and JavaScript (37.19\%).
In contrast, it encounters more significant challenges in C++ (14.91\%) and C (19.67\%), while producing moderate outcomes in Python (20.34\%), C\# (31.91\%), and Java (34.99\%).
Overall, AVR techniques and PLM approaches perform the best performance in Go with a median of 20.96\%, followed by Java at 14.58\%, JavaScript at 9.71\%, C\# at 8.51\%, C at 8.20\%, Python at 7.34\% and C++ at 3.64\%. 

Based on the above data analysis, we observe that Encoder-decoder PLMs perform best among three types of PLM techniques and VulMaster performs best on all seven metrics among all AVRs and PLMs.
Given that the backbone model of VulMaster is CodeT5 which belongs to Encoder-decoder PLMs and its input incorporates code AST information and CWE information, researchers would consider utilizing Encoder-decoder architecture and more information about code and CWE knowledge for further improving learning-based automated vulnerability repair in the multilingual vulnerability context.
Using Fusion-in-Decoder architecture, the input length of VulMaster which is 5,120 tokens is significantly broadened compared with the input length of the original CodeT5 model which is 512 tokens.
Therefore, a model with a larger input length or approach that would broaden the model input length is one direction that researchers could take into consideration.

\begin{tcolorbox}\textbf{RQ1 Summary:}
VulMaster and CodeT5 achieve the best performance among AVR and PLM techniques, with 28.94\% and 14.33\% in the EM with beam 1 metric, respectively.
VulMaster is the best-performing technique across all seven metrics, surpassing all other methods by a margin of 82.82\% to 628.97\% on the EM with beam 1 metric.
All techniques demonstrate optimal performance with Go while showing their relatively poorest results with C/C++.
\end{tcolorbox}

\subsection{RQ2: What is the performance of state-of-the-art LLMs in repairing multilingual vulnerability?}
\label{subsec:RQ2}
\noindent
\textbf{\emph{\underline{Approach.}}}
This research question focuses on comparing language-agnostic LLMs and various learning strategies. We investigated five state-of-the-art LLMs and four prompting strategies to assess their effectiveness. 
Prior work~\cite{10479409} showed that LLMs perform poorly in automated vulnerability repair when using the special tokens as a part of the prompt.
Additionally, closed-source LLMs accessed via API cannot accept special tokens.
Thus, to improve the prompt effectiveness and maintain consistency with open-source LLMs, we applied the Sequence-to-Sequence paradigm for all the prompting strategies (including both zero-shot and few-shot) and inference, where the input is a vulnerable function and the expected output is a corresponding repaired function.
Moreover, we approach multilingual vulnerability repair as a multi-task learning problem, where repairing vulnerabilities in each programming language represents a distinct task. 
Drawing from previous research~\cite{Chung2022ScalingIL, Mueller2024MultiTaskTM}, we applied instruction tuning as a multi-task learning strategy to empower LLMs to transfer knowledge across programming languages. 
On that basis, we can further apply language-specific few-shot prompts to enhance the in-context learning for multilingual vulnerability repair.
Due to varying prompting strategies, we restricted the vulnerable function length to 1024 tokens.
Section~\ref{subsec:models} and Section~\ref{subsec:strategy} present detailed information about the LLMs and their strategies. 
Below, we particularly describe the instruction tuning process.

Our study aims to develop a language-agnostic paradigm for repairing software vulnerabilities using LLMs.
To adapt these models to multilingual vulnerability repair, we employed instruction tuning on our complete training dataset of 7,448 function pairs spanning seven programming languages.
In the first step, we applied zero-shot prompting to convert the multilingual training set into an instruction-filled fine-tuning set, consisting of $<$\textit{structured instruction}, \textit{vulnerable code}, \textit{repaired code}$>$.
Next, we applied Low-Rank Adaption (LoRA) to a subset of weight matrices in the LLM, adapting only the attention weights for multilingual vulnerability repair while freezing the multilayer perceptron (MLP) modules to reduce trainable parameters.
We then trained the LoRA-based LLM on the instruction-filled dataset using supervised learning. 
After we obtained the instruction-tuned LLMs, we used both zero-shot and few-shot prompting strategies (as described in Section~\ref{subsec:strategy}) to generate the repaired functions. 
For example, in the few-shot prompting strategy, we used BM25 to retrieve three language-specific pairs of vulnerable and repaired functions as examples, then prompted the instruction-tuned LLM to generate the fixes.

To evaluate the effectiveness of the studied LLMs and their strategies, we used the similar metrics as in RQ1: Exact Match (EM) with beam size 1, BLEU score, and ROUGE-1, ROUGE-2, and ROUGE-L metrics. 
We limit our analysis to beam size of 1 to maintain consistency between the open-source and closed-source models, as closed-source models do not support beam search.
In addition, we performed a McNemar test~\cite{mccrum2008correct}, a non-parametric statistical method, to assess the statistical significance of differences between the best-performing LLM-based approach and VulMaster (the best-performing existing AVR technique) in terms of EM with beam 1.

\begin{table}[t]
    \caption{Performance of LLMs and their strategies on multilingual vulnerability repair}
    \label{tab:rq2}
    \centering
    \tabcolsep=3.0mm
    \small
    \begin{adjustbox}{max width=1.0\textwidth, center}
    \begin{tabular}{lrrrrr}
        \toprule
        \toprule
        \textbf{Technique} & \textbf{EM with beam 1} & \textbf{BLEU-4} & \textbf{ROUGE-1} & \textbf{ROUGE-2} & \textbf{ROUGE-L} \\ 
        \midrule
        \rowcolor{mygray}\multicolumn{6}{l}{\textbf{Zero-shot prompting}} \\ 
        \midrule
            DeepSeek-Coder & 0.09\% & 0.0489 & 0.2725 & 0.1354 & 0.2708 \\
            Code Llama      & 0.37\% & 0.0606 & 0.3893 & 0.2007 & 0.3864 \\
            Llama 3 & 0.19\% & 0.0445 & 0.3189 & 0.1583 & 0.3173 \\
            GPT-3.5-Turbo  & 0.23\% & 0.0588 & 0.4071 & 0.2118 & 0.4055 \\
            GPT-4o         & 0.33\% & 0.0806 & 0.4084 & 0.2188 & 0.4057 \\
        \midrule
        \rowcolor{mygray}\multicolumn{6}{l}{\textbf{Few-shot prompting}} \\ 
        \midrule
            DeepSeek-Coder & 14.15\% & 0.3857 & 0.6107 & 0.5192 & 0.6079 \\
            Code Llama      & 2.52\%  & 0.0847 & 0.2859 & 0.1645 & 0.2825 \\
            Llama 3          & 11.48\% & 0.1574 & 0.4678 & 0.3491 & 0.4658 \\
            GPT-3.5-Turbo  & 17.32\% & 0.3503 & 0.6595 & 0.5474 & 0.6577 \\
            GPT-4o         & 26.89\% & 0.6193 & 0.7941 & 0.7243 & 0.7920 \\
        \midrule
        \rowcolor{mygray}\multicolumn{6}{l}{\textbf{Instruction-tuning + Zero-shot prompting}} \\ 
        \midrule
            DeepSeek-Coder & 0.42\% & 0.7240 & 0.8585 & 0.8143 & 0.8553 \\
            Code Llama      & 3.08\% & 0.8361 & 0.9183 & 0.8840 & 0.9158 \\
            Llama 3 & 3.36\% & 0.4914 & 0.8274 & 0.7446 & 0.8245 \\
            GPT-3.5-Turbo  & 5.79\% & 0.7067 & 0.8419 & 0.7865 & 0.8373 \\
            GPT-4o         & 7.24\% & 0.7729 & 0.8865 & 0.8408 & 0.8833 \\
        \midrule
        \rowcolor{mygray}\multicolumn{6}{l}{\textbf{Instruction-tuning + Few-shot prompting}} \\ 
        \midrule
           DeepSeek-Coder & 20.96\% & 0.7579 & 0.8553 & 0.8127 & 0.8502 \\
            Code Llama      & 18.25\% & 0.7770 & 0.8557 & 0.8128 & 0.8498 \\
            Llama 3 & 18.30\% & 0.3496 & 0.6513 & 0.5819 & 0.6478 \\
            GPT-3.5-Turbo  & 16.34\% & 0.7617 & 0.8701 & 0.8255 & 0.8662 \\
            GPT-4o         & \textbf{28.71\%} & \textbf{0.8448}& \textbf{0.9232} & \textbf{0.8936} & \textbf{0.9202} \\
        \bottomrule
        \bottomrule
    \end{tabular}

    \end{adjustbox}
\end{table}

\begin{figure*}[]
    \centering
    \includegraphics[width=1.\linewidth]{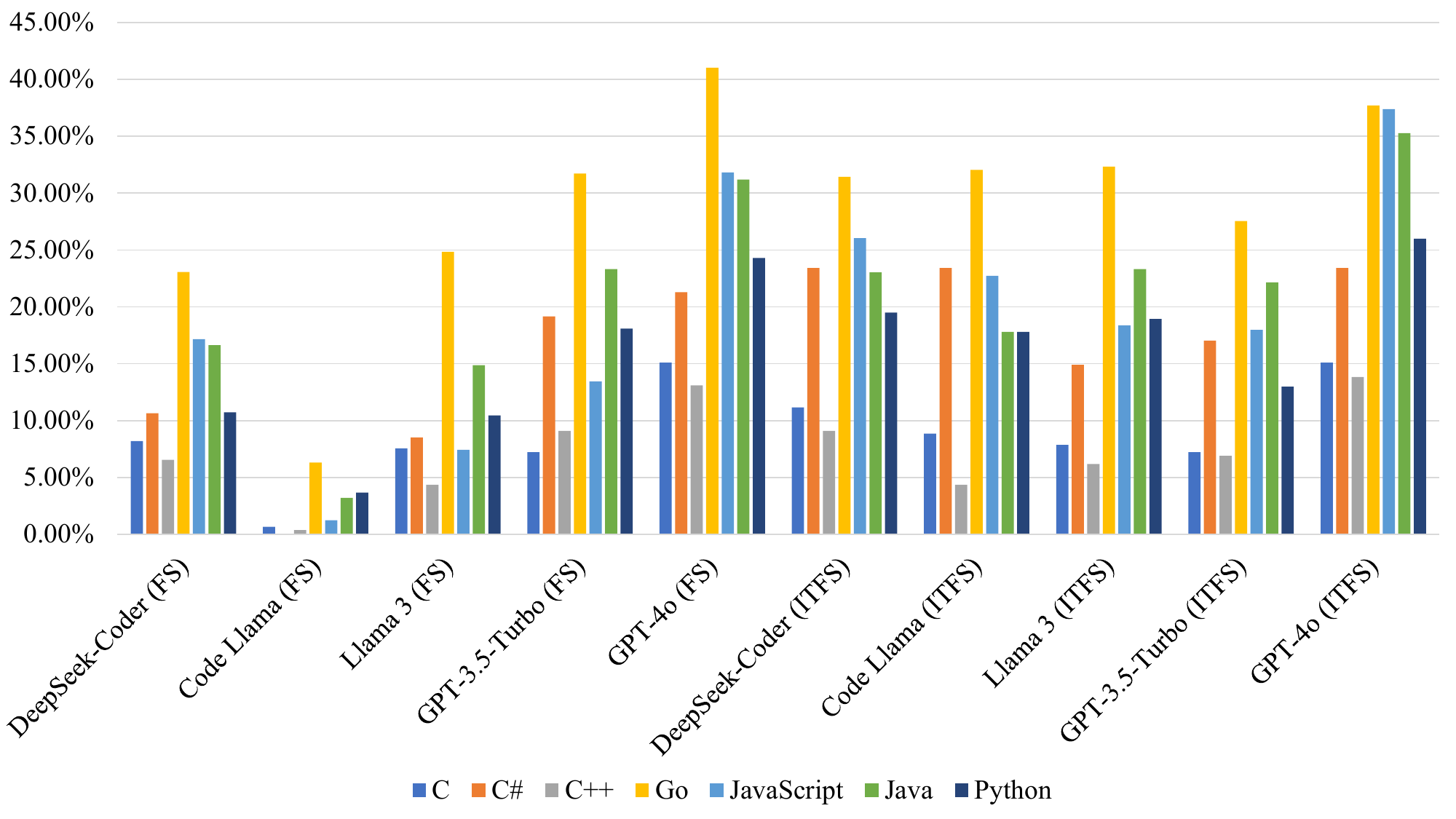}
    \caption{Performance of LLM techniques across seven programming languages (y-axis: Exact Match score with beam size 1, FS: few-shot prompting, and ITFS: instruction-tuning with few-shot prompting)}.
    \label{fig:rq2}
\end{figure*}

\begin{figure*}[]
    \centering
    \includegraphics[width=.7\linewidth]{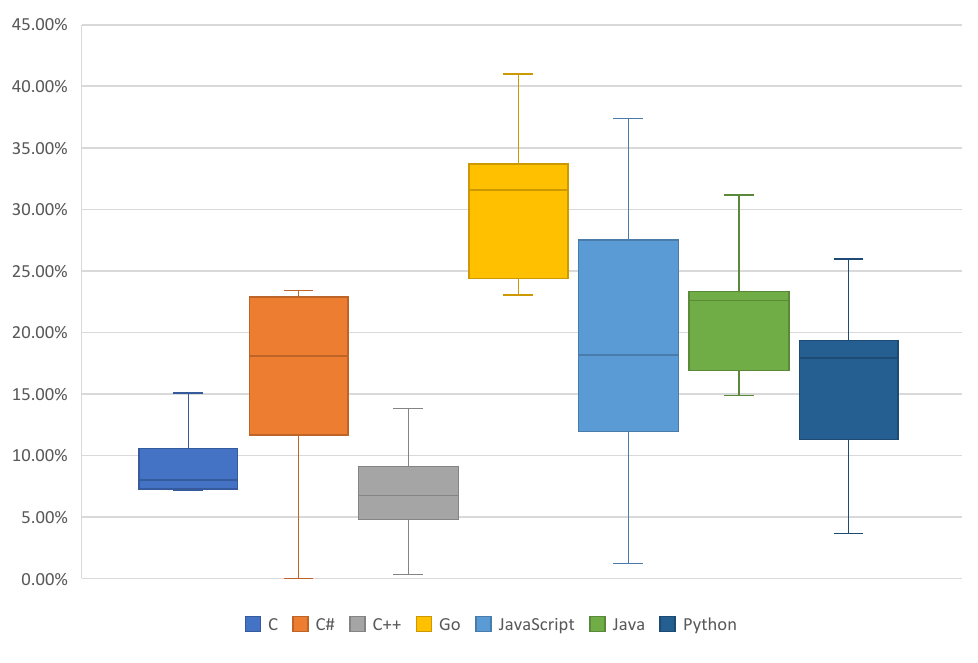}
    \caption{Overall performance of LLM techniques across seven programming languages (y-axis: Exact Match score with beam size 1)}
    \label{fig:rq2_box}
\end{figure*}

\noindent
\textbf{\emph{\underline{Results.}}} 
Table~\ref{tab:rq2} illustrates the performance of five LLMs under four different strategies in terms of EM, BLEU, and ROUGE scores in the multilingual vulnerability repair context. 
Within each metric, the values in bold indicate the technique that exhibits the best performance among all LLMs with different strategies.
Figure~\ref{fig:rq2} and Figure~\ref{fig:rq2_box} present the performance of LLMs of the few-shot prompting strategy and instruction-tuning with few-shot prompting strategy across seven programming languages in terms of EM with beam size 1.

First, under the zero-shot prompting strategy, both open-source and closed-source LLMs exhibit poor performance on this multilingual vulnerability repair task.
Nevertheless, the performance of LLMs is extremely improved by repairing vulnerable code with the few-shot prompting strategy which uses the BM25 algorithm to retrieve three examples from the training dataset with the highest similarity.
Specifically, among all LLMs evaluated, Code Llama achieves an EM of 0.37\% using a beam size of 1 under zero-shot prompting conditions, which modestly increases to 2.52\% with few-shot prompting, marking the lowest relative gain at 581.08\%.
In contrast, DeepSeek-Coder begins with an EM of 0.09\% under zero-shot conditions, which dramatically escalates to 14.15\% with few-shot prompting, resulting in the highest relative improvement recorded at 15,622.22\% in terms of EM with beam 1.
This shows the strong power of prompt engineering and implies that providing more information to LLMs could get more valuable results.

Second, instruction-tuning boosts the performance of the majority of models in terms of EM with beam 1 and also improves some models in terms of BLEU and ROUGE metrics.
For example, instruction-tuning Code Llama with few-shot prompting outperforms Code Llama with few-shot prompting with an improvement of 624.21\% in terms of EM with beam 1.
This suggests that the instruction-tuning strategy can substantially enhance the performance of LLMs in multilingual vulnerability repair.
Additionally, upon inspecting their code generation metric results, we observed that despite the small instruction-tuning epoch due to limited resources, few LLMs have learned the repair pattern for multilingual vulnerability repair to some extent.

Additionally, among the five models employing four prompting strategies, the GPT-4o model with instruction-tuning and few-shot prompting achieves the best performance.
It records scores of 28.71\% in EM with beam 1, 0.8448 in BLEU-4, 0.9232 in ROUGE-1, 0.8936 in ROUGE-2, and 0.9202 in ROUGE-L.
Moreover, Table~\ref{tab:4_rq3_mcnemar} presents the results of the McNemar test conducted to compare the performance of the instruction-tuning GPT-4o model with few-shot prompting strategy and VulMaster.
The p-value obtained from the McNemar test is 0.8314, which is well above the conventional significance threshold of 0.05. 
This indicates that there is no statistically significant difference between the classification results of GPT-4o and VulMaster. 
Additionally, the odds ratio (OR) is reported as 0.9721. 
This value, close to 1, suggests that the difference in classification errors between the two models is minimal, with neither model demonstrating a notable advantage over the other. 
In conclusion, results of the McNemar test suggest that there is no significant difference in performance between GPT-4o and VulMaster in multilingual vulnerability repair, which exhibits the strong power of instruction-tuning with few-shot prompting strategy.

\begin{table}[htbp]
    \centering
    \caption{McNemar test results between GPT-4o and VulMaster}
    \label{tab:4_rq3_mcnemar}
    \begin{tabular}{|c|c|c|}
        \hline
        \textbf{Approach} & \textbf{McNemar.p} & \textbf{McNemar.OR} \\ \hline
         GPT-4o vs. VulMaster & 0.8314 & 0.9721 \\ \hline
    \end{tabular}
    
\end{table}

In Figure \ref{fig:rq2} and Figure ~\ref{fig:rq2_box}, owing to the subpar performance observed with zero-shot prompting and instruction-tuning with zero-shot prompting strategies, we present only the results from models utilizing few-shot prompting and instruction-tuning with few-shot prompting strategies. 
Consistent with the findings from RQ1, the LLMs demonstrate comparable performance across different programming languages, with the highest performance noted in Go and the lowest in C/C++.
Instruction-tuning GPT-4o with few-shot prompting shows its best performance in C, C\#, C++, JavaScript, Java, and Python, achieving EM with beam 1 scores of 15.08\%, 23.40\%, 13.82\%, 37.40\%, 35.28\%, and 25.99\%, respectively.
Overall, LLM techniques perform best in Go with a median of 31.59\%, followed by 
Java at 22.59\%, JavaScript at 18.18\%, C\# at 18.09\%, Python at 17.94\%, C at 8.03\% and C++ at 6.73\%. 
Notably, instruction-tuning of both DeepSeek-Coder and Code Llama with few-shot prompting achieved the same results as GPT-4o on C\#, indicating the effectiveness of instruction-tuning in enhancing the capabilities of LLMs.
Interestingly, GPT-4o with few-shot prompting outperformed its instruction-tuned counterpart in Go, with an EM with beam 1 score of 41.02\%.
This could be attributed to the limited scale of the existing fine-tuning datasets, which may not fully unleash the potential of GPT-4o.

\begin{tcolorbox}\textbf{RQ2 Summary:}
Instruction-tuning with few-shot prompting strategy is the best technique among all strategies for LLMs on multilingual vulnerability repair and instruction-tuning GPT-4o with few-shot prompting strategy achieves the best performance of 28.71\% in terms of EM with beam 1.
The performances of LLMs vary from different programming languages and LLMs perform best on Go and worst on C/C++, which is similar to AVR and PLM techniques.

\end{tcolorbox}

\subsection{RQ3: What are the strengths and weaknesses of the existing automated vulnerability techniques?}
\label{subsec:RQ3}
\noindent
\textbf{\emph{\underline{Approach.}}} 
This research question aims to gain an understanding of the performance characteristics of various AVR, PLM techniques, and LLM approaches with different strategies.
Therefore, following the experimental design of RQ1 and RQ2, we conducted a comprehensive analysis to explore the degree of their orthogonality from two different perspectives:
\begin{itemize}[leftmargin=10pt]
    \item \textbf{Between AVR techniques.} 
    Based on the findings of RQ1 and RQ2, we selected the best-performing AVR technique, PLM technique, and LLM technique.
    Specifically, the selected techniques are VulMaster for AVR techniques, CodeT5 for three PLM architectures, and instruction-tuning GPT-4o with few-shot prompting for LLM techniques.

    \item \textbf{Between LLM strategies.} 
    Based on the findings of RQ2, we selected the best-performing LLM techniques within four LLM strategies, i.e., zero-shot prompting (ZS), few-shot prompting (FS), instruction-tuning with zero-shot prompting (ITZS), and instruction-tuning with few-shot prompting strategies (ITFS).
    Specifically, the selected techniques are Code Llama with zero-shot prompting, GPT-4o with few-shot prompting, instruction-tuning GPT-4o with zero-shot prompting, and instruction-tuning GPT-4o with few-shot prompting, each representing their respective LLM strategies.
\end{itemize}

Based on the above two perspectives, we further conducted a two-level analysis: (1) the unique correct/incorrect repairs, and (2) the repair performance across the dangerous CWE-IDs.
Regarding the first level, we employed Venn diagrams to assess the unique correct/incorrect repairs across various studied AVR techniques.
Regarding the second level, we investigated the repair performance of studied techniques using testing data based on the 2023 CWE Top 25 Most Dangerous Software Weaknesses, as published by the CWE community.\footnote{\url{https://cwe.mitre.org/top25/archive/2023/2023_top25_list.html}}
Among the top 25 dangerous CWE-IDs, all of them are involved in our testing data.
Finally, we investigate the highest and lowest CWEs in terms of the average EM scores with beam 1, across all AVR, PLM, and LLM techniques for seven programming languages.

\begin{figure*}[]
    \centering
    \includegraphics[width=.8\linewidth]{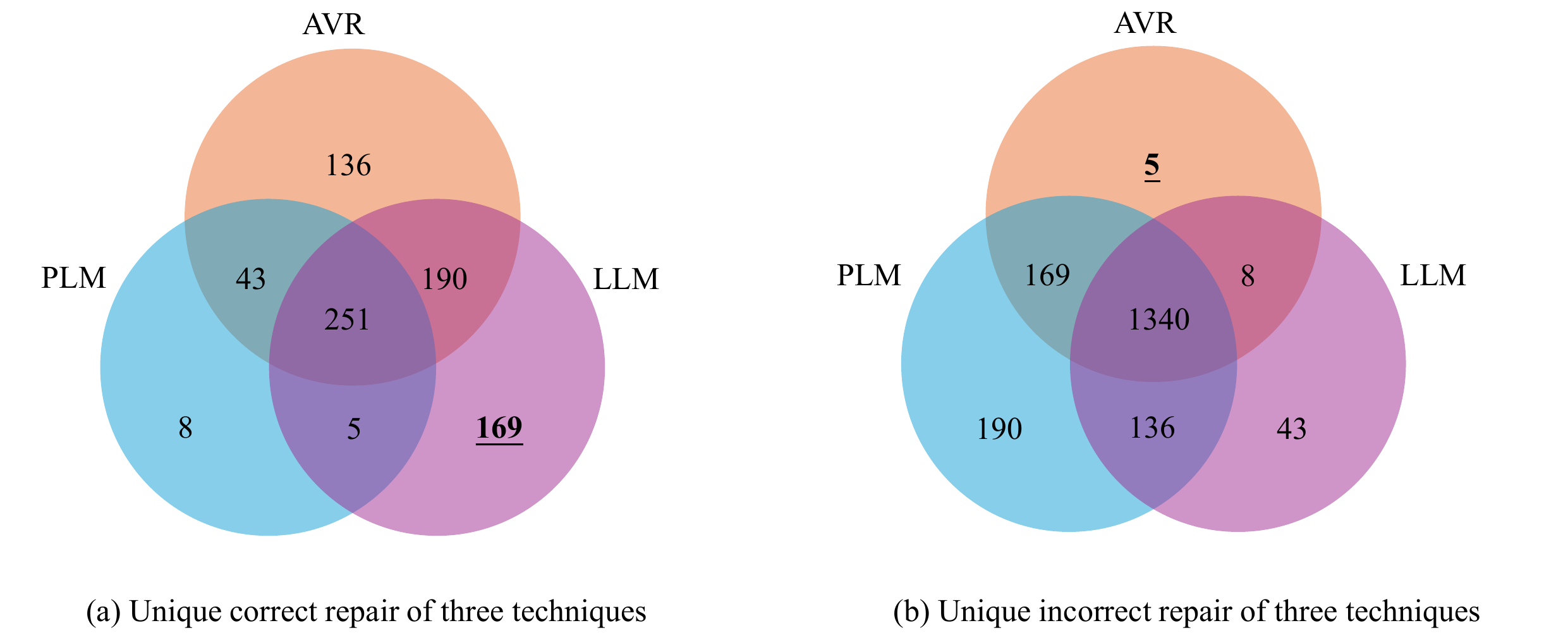}
    \caption{Unique correct repair and unique incorrect repair across three representative techniques}
    \label{fig:rq3_venn}
\end{figure*}

\begin{figure*}[]
    \centering
    \includegraphics[width=\linewidth]{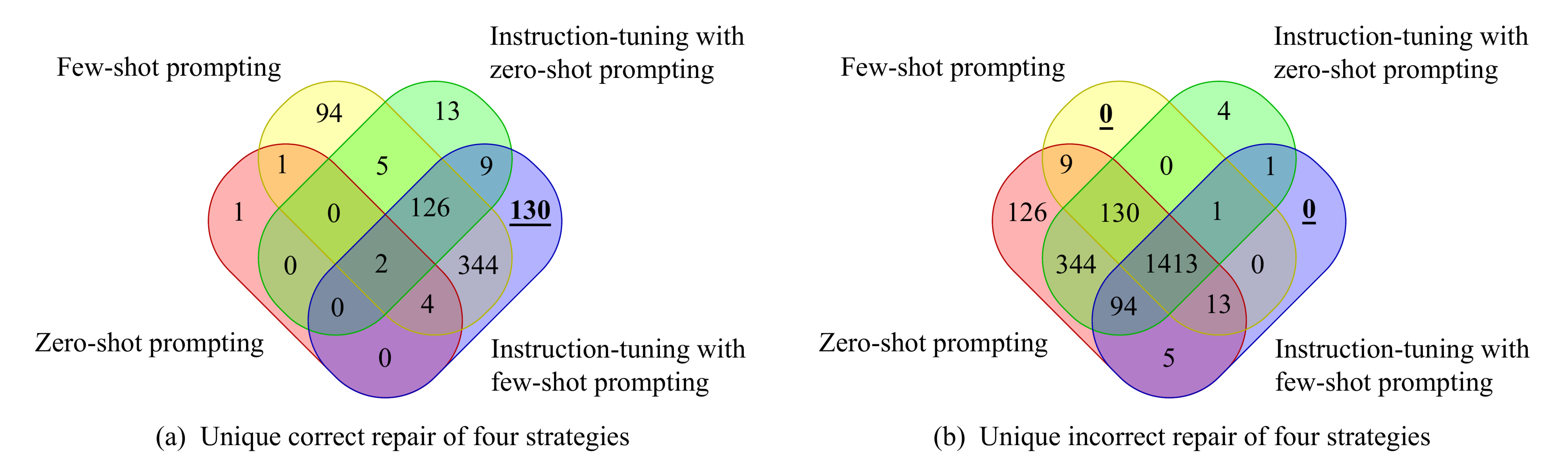}
    \caption{Unique correct repair and unique incorrect repair across four LLM strategies}
    \label{fig:rq3_venn_llm}
\end{figure*}

\noindent
\textbf{\emph{\underline{Results.}}} 
Figure~\ref{fig:rq3_venn} and Figure~\ref{fig:rq3_venn_llm} present the Venn diagrams that demonstrate the intersection of correct/incorrect repairs among the studied AVR techniques based on two analyzed perspectives (i.e., AVR techniques and LLM strategies) in terms of EM with beam 1. 
Overlap areas denote shared correct/incorrect repairs among multiple AVR techniques, while non-overlapping areas signify the unique correct/incorrect repairs of each AVR technique. 
Table~\ref{tab:rq3_top25} and Table~\ref{tab:rq3_top25_llm} further show the repair performance of studied AVR techniques and the LLM strategies on the top 25 most dangerous CWE-IDs in 2023 in terms of EM with beam 1 and the values highlighted in bold denote the technique or strategy that exhibits the best performance for corresponding CWE-ID.
Table~\ref{tab:rq3_average_em} further displays the highest and lowest CWEs in terms of the average EM scores with beam 1, across all AVR, PLM, and LLM techniques for seven programming languages.
Table~\ref{tab:deep_analysis} provides an analysis of the best and worst performing CWEs in relation to the most effective LLM and AVR techniques.

\textbf{Between AVR techniques.}
From Figure ~\ref{fig:rq3_venn}a, we find that the LLM technique achieves the best performance compared to AVR and PLM in terms of the unique correct repairs.
Specifically, the LLM technique demonstrates 169 unique correct repairs, significantly outperforming AVR, which achieves 136 unique correct repairs (a 24.26\% improvement), and PLM, which only has 8 unique correct repairs (a 2012.50\% improvement).
From Figure ~\ref{fig:rq3_venn}b, we find that the LLM technique also outperforms PLM in terms of unique incorrect repairs.
It only has 43 unique incorrect repairs, which is significantly fewer than PLM (190).
These results demonstrate the effectiveness of instruction-tuning with few-shot prompting strategy for LLMs in multilingual vulnerability repair.

From Table ~\ref{tab:rq3_top25}, we observe that the LLM technique outperforms the other two categories, AVR and PLM.
On average, LLM successfully repairs 32.48\% of the vulnerable functions (1204) associated with the top 25 most dangerous CWE-IDs in 2023, surpassing the performance of AVR (31.15\%) and PLM (17.44\%).
Notably, the LLM technique achieves the best performance on 17 out of the 25 CWE-IDs, compared to 14 for AVR and none for PLM.
These findings underscore the efficacy of LLM in addressing dangerous vulnerabilities, thereby reinforcing the results presented in RQ2.

\begin{table*}[t]
    \caption{The Exact Match score with beam size 1 of three techniques across the Top 25 Most Dangerous CWE-IDs in 2023}
    \label{tab:rq3_top25}
    \centering
    \small
    \begin{adjustbox}{max width=1.0\textwidth, center}
    \begin{tabular}{ccrrrc}
        \toprule
        \toprule
        Rank & CWE-ID & PLM & AVR & LLM & Total \\
        \midrule
        1 & CWE-787 (Out-of-bounds Write) & 11(14.47\%)  & \textbf{12(15.79\%)}  & \textbf{12(15.79\%)}  & 76   \\
        2 & CWE-79 (Cross-site Scripting) & 53(21.03\%)  & \textbf{106(42.06\%)} & 103(40.87\%) & 252  \\
        3 & CWE-89 (SQL Injection)& 6(9.38\%)    & 14(21.88\%)  & \textbf{20(31.25\%) } & 64   \\
        4 & CWE-416 (Use After Free) & 4(13.33\%)   & \textbf{8(26.67\%)}  & \textbf{8(26.67\%)}  & 30   \\
        5 & CWE-78 (OS Command Injection) & 2(7.41\%)    & \textbf{6(22.22\%)}  & 5(18.52\%)   & 27   \\
        6 & CWE-20 (Improper Input Validation) & 11(10.78\%)  & 27(26.47\%)  & \textbf{28(27.45\%)}  & 102  \\
        7 & CWE-125 (Out-of-bounds Read) & 0(0.00\%)    & \textbf{2(4.55\%)}    & 0(0.00\%)    & 44   \\
        8 & CWE-22 (Path Traversal) & 1(1.18\%)    & 11(12.94\%)  & \textbf{17(20.00\%)}  & 85   \\
        9 & CWE-352 (Cross-Site Request Forgery) & 12(23.53\%)  & \textbf{18(35.29\%)}  & 16(31.37\%)  & 51   \\
        10 & CWE-434 (Unrestricted Upload of File with Dangerous Type) & 0(0.00\%)    & \textbf{1(20.00\%)}   & \textbf{1(20.00\%)}  & 5    \\
        11 & CWE-862 (Missing Authorization) & 0(0.00\%)    & 0(0.00\%)    & 0(0.00\%)    & 4    \\
        12 & CWE-476 (NULL Pointer Dereference) & 5(7.69\%)    & 12(18.46\%)  & \textbf{14(21.54\%)} & 65   \\
        13 & CWE-287 (Improper Authentication) & 82(63.57\%)  & \textbf{93(72.09\%)}  & \textbf{93(72.09\%)}  & 129  \\
        14 & CWE-190 (Integer Overflow or Wraparound) & 4(8.16\%)    & \textbf{10(20.41\%)}  & 6(12.24\%)   & 49   \\
        15 & CWE-502 (Deserialization of Untrusted Data) & 5(18.52\%)   & \textbf{11(40.74\%)}  & 8(29.63\%)   & 27   \\
        16 & CWE-77 (Command Injection) & 3(7.89\%)    & 8(21.05\%)   & \textbf{18(47.37\%)}  & 38   \\
        17 & CWE-119 (Improper Restriction of Operations within the Bounds of a Memory Buffer) & 2(7.14\%) & 4(14.29\%) & \textbf{5(17.86\%)} & 28 \\
        18 & CWE-798 (Use of Hard-coded Credentials) & 0(0.00\%)    & 0(0.00\%)    & \textbf{1(50.00\%)}   & 2    \\
        19 & CWE-918 (Server-Side Request Forgery) & 2(6.25\%)    & \textbf{7(21.88\%)}   & \textbf{7(21.88\%)}   & 32   \\
        20 & CWE-306 (Missing Authentication for Critical Function)    & 2(18.18\%)   & \textbf{7(63.64\%)}  & \textbf{7(63.64\%)}   & 11   \\
        21 & CWE-362 (Race Condition) & 0(0.00\%)    & \textbf{5(45.45\%)}   & \textbf{5(45.45\%)}  & 11   \\
        22 & CWE-269 (Improper Privilege Management) & 0(0.00\%)    & 3(23.08\%)   & \textbf{4(30.77\%)}   & 13   \\
        23 & CWE-94 (Code Injection) & 2(8.70\%)    & 3(13.04\%)   & \textbf{6(26.09\%)}   & 23   \\
        24 & CWE-863 (Incorrect Authorization) & 3(9.09\%)    & \textbf{7(21.21\%)}   & \textbf{7(21.21\%)}   & 33   \\
        25 & CWE-276 (Incorrect Default Permissions) & 0(0.00\%)    & 0(0.00\%)    & 0(0.00\%)    & 3    \\
        \midrule
         & Average & 210(17.44\%) & 375(31.15\%) & \textbf{391(32.48\%)} & 1204 \\
        \bottomrule
        \bottomrule
    \end{tabular}
    \end{adjustbox}
    \begin{tablenotes}
    \footnotesize
        \item[*] The number in the parentheses represent the corresponding accuracy of techniques and Total column represents the total amount of data for corresponding CWE-ID in our test dataset.
    \end{tablenotes}
\end{table*}

\textbf{Between LLM strategies.}
From Figure ~\ref{fig:rq3_venn_llm}a, we observe that the ITFS technique achieves the best performance in terms of unique correct repairs when compared to ZS, FS, and ITZS. 
The ITFS technique achieves 130 unique correct repairs, which represents a substantial improvement over the other strategies. 
Specifically, it outperforms ZS, which achieves only 1 unique correct repair, resulting in an improvement of 12,900.00\%. 
It also surpasses ITZS, which yields 13 unique correct repairs, with an improvement of 900.00\%, and FS, which produces 94 unique correct repairs, showing an improvement of 38.30\%.
Moreover, as illustrated in Figure ~\ref{fig:rq3_venn_llm}b, the LLM technique also excels in avoiding unique incorrect repairs, with no such cases identified—significantly fewer than ZS (126) and ITZS (4).
These results emphasize the effectiveness of instruction-tuning with few-shot prompting strategy in enhancing LLM performance for multilingual vulnerability repair.

Table ~\ref{tab:rq3_top25_llm} provides additional evidence of the superior performance of ITFS compared to the other three LLM techniques.
On average, the ITFS approach successfully repairs vulnerabilities in 32.48\% of the functions associated with the top 25 most dangerous CWE-IDs, demonstrating a significant improvement over other methods.
Specifically, it outperforms ITZS, which repairs only 10.55\% of these vulnerable functions, achieving a 207.87\% relative improvement.
It also surpasses FS, which repairs 29.65\% of the vulnerabilities, reflecting a 9.52\% enhancement in effectiveness.
In comparison to ZS, which addresses only 0.50\% of the vulnerable functions, ITFS achieves a striking improvement of 6416.67\%, vastly outperforming ZS.
Notably, ITFS achieves the best performance on 18 out of the 25 CWE-IDs, compared to 11 for FS and none for ITZS and ZS.
These findings further underscore the efficacy of instruction tuning with few-shot prompting in significantly improving LLM capabilities for addressing dangerous vulnerabilities, reinforcing the results presented in RQ2.

\begin{table*}[t]
    \caption{The Exact Match score with beam size 1 of four LLM strategies across the Top 25 Most Dangerous CWE-IDs in 2023}
    \label{tab:rq3_top25_llm}
    \centering
    \small
    \begin{adjustbox}{max width=1.0\textwidth, center}
    
    \begin{tabular}{ccrrrrc}
        \toprule
        \toprule
        Rank & ID & ZS & FS & ITZS & ITFS & total \\
    \midrule 
    1 & CWE-787 (Out-of-bounds Write)               & 0(0.00\%) & \textbf{13(17.11\%)}  & 5(6.58\%)    & 12(15.79\%)  & 76   \\
    2 & CWE-79 (Cross-site Scripting)               & 0(0.00\%) & 87(34.52\%)  & 32(12.70\%)  & \textbf{103(40.87\%)} & 252  \\
    3 & CWE-89 (SQL Injection)   & 1(1.56\%) & 17(26.56\%)  & 1(1.56\%)    & \textbf{20(31.25\%)}  & 64   \\
    4 & CWE-416 (Use After Free) & 0(0.00\%) & 6(20.00\%)   & 0(0.00\%)    & \textbf{8(26.67\%)}   & 30   \\
    5 & CWE-78 (OS Command Injection)               & 0(0.00\%) & 4(14.81\%)   & 0(0.00\%)    & \textbf{5(18.52\%)}   & 27   \\
    6 & CWE-20 (Improper Input Validation)          & 0(0.00\%) & \textbf{28(27.45\%)}  & 4(3.92\%)    & \textbf{28(27.45\%)}  & 102  \\
    7 & CWE-125 (Out-of-bounds Read)                & 0(0.00\%) & \textbf{1(2.27\%)}    & 0(0.00\%)    & 0(0.00\%)    & 44   \\
    8 & CWE-22 (Path Traversal)  & 0(0.00\%) & 12(14.12\%)  & 0(0.00\%)    & \textbf{17(20.00\%)}  & 85   \\
    9 & CWE-352 (Cross-Site Request Forgery)        & 0(0.00\%) & \textbf{17(33.33\%)}  & 3(5.88\%)    & 16(31.37\%)  & 51   \\
    10 & CWE-434 (Unrestricted Upload of File with Dangerous Type) & 0(0.00\%) & \textbf{1(20.00\%)} &  0(0.00\%) & \textbf{1(20.00\%)} & 5 \\
    11 & CWE-862 (Missing Authorization)             & 0(0.00\%) & 0(0.00\%)    & 0(0.00\%)    & 0(0.00\%)    & 4    \\
    12 & CWE-476 (NULL Pointer Dereference)          & 2(3.08\%) & \textbf{14(21.54\%)}  & 4(6.15\%)    & \textbf{14(21.54\%)}  & 65   \\
    13 & CWE-287 (Improper Authentication)           & 1(0.78\%) & 90(69.77\%)  & 74(57.36\%)  & \textbf{93(72.09\%)}  & 129  \\
    14 & CWE-190 (Integer Overflow or Wraparound)    & 0(0.00\%) & \textbf{10(20.41\%)}  & 0(0.00\%)    & 6(12.24\%)   & 49   \\
    15 & CWE-502 (Deserialization of Untrusted Data) & 0(0.00\%) & \textbf{8(29.63\%)}   & 0(0.00\%)    & \textbf{8(29.63\%)}   & 27   \\
    16 & CWE-77 (Command Injection)                  & 1(2.63\%) & 15(39.47\%)  & 3(7.89\%)    & \textbf{18(47.37\%)}  & 38   \\
    17 & CWE-119 (Improper Restriction of Operations within the Bounds   of a Memory Buffer) & 0(0.00\%) & 3(10.71\%) & 1(3.57\%) & \textbf{5(17.86\%)} & 28 \\
    18 & CWE-798 (Use of Hard-coded Credentials)     & 0(0.00\%) & 0(0.00\%)    & 0(0.00\%)    & \textbf{1(50.00\%)}   & 2    \\
    19 & CWE-918 (Server-Side Request Forgery)       & 0(0.00\%) & 5(15.63\%)   & 0(0.00\%)    & \textbf{7(21.88\%)}   & 32   \\
    20 & CWE-306 (Missing Authentication for Critical Function) & 0(0.00\%) & \textbf{7(63.64\%)} & 0(0.00\%) & \textbf{7(63.64\%)} & 11 \\
    21 & CWE-362 (Race Condition) & 0(0.00\%) & \textbf{5(45.45\%)}   & 0(0.00\%)    & \textbf{5(45.45\%)}   & 11   \\
    22 & CWE-269 (Improper Privilege Management)     & 0(0.00\%) & \textbf{5(38.46\%)}   & 0(0.00\%)    & 4(30.77\%)   & 13   \\
    23 & CWE-94 (Code Injection)  & 1(4.35\%) & 4(17.39\%)   & 0(0.00\%)    & \textbf{6(26.09\%)}   & 23   \\
    24 & CWE-863 (Incorrect Authorization)           & 0(0.00\%) & 5(15.15\%)   & 0(0.00\%)    & \textbf{7(21.21\%)}   & 33   \\
    25 & CWE-276 (Incorrect Default Permissions)     & 0(0.00\%) & 0(0.00\%)    & 0(0.00\%)    & 0(0.00\%)    & 3    \\
    \midrule
  & Average & 6(0.50\%) & 357(29.65\%) & 127(10.55\%) & \textbf{391(32.48\%)} & 1204 \\
        \bottomrule
        \bottomrule
    \end{tabular}
    \end{adjustbox}
    \begin{tablenotes}
    \footnotesize
        \item[*] The number in the parentheses represent the corresponding accuracy of techniques and Total column represents the total amount of data for corresponding CWE-ID in our test dataset.
    \end{tablenotes}
\end{table*}

From Table ~\ref{tab:rq3_average_em}, we can see that the performance of the techniques varies across different languages.
For instance, the language C exhibits the highest repair rate for CWE-327 (Use of a Broken or Risky Cryptographic Algorithm) with an EM of 31.25\%, yet fails to repair CWE-121 (Stack-based Buffer Overflow), scoring 0.00\%.
Similarly, C\# shows a considerable ability to repair CWE-918 (Server-Side Request Forgery) with an EM of 23.13\%, but is ineffective at repairing CWE-347 (Improper Verification of Cryptographic Signature), which also scores 0.00\%.
Notably, Go achieves the highest EM among the languages with a 60.25\% repair rate for CWE-287 (Improper Authentication), contrasting with its inability to repair CWE-208 (Observable Timing Discrepancy).
This result underscores the varied effectiveness of vulnerability repair techniques across different programming environments and specific vulnerabilities, pointing to potential areas for further refinement in repair capabilities.

\begin{table*}[t]
    \caption{Highest and lowest average Exact Match score with beam size 1 of all techniques across seven programming languages}
    \label{tab:rq3_average_em}
    \centering
    \small
    \begin{adjustbox}{max width=1.0\textwidth, center}
    
    \begin{tabular}{ccrcr}
        \toprule
        Language   & Highest EM CWE                                            & EM with beam 1 & Lowest EM CWE                                    & EM with beam 1 \\ 
        \midrule
        C & CWE-327: Use of a Broken or Risky Cryptographic Algorithm & 31.25\%        & CWE-121: Stack-based Buffer Overflow             & 0.00\%         \\
        C\# & CWE-918: Server-Side Request Forgery (SSRF) & 23.13\% & CWE-347: Improper Verification of Cryptographic Signature           & 0.00\% \\
        C++ & CWE-191: Integer Underflow (Wrap or Wraparound) & 37.50\% & CWE-116: Improper Encoding or Escaping of Output & 0.00\% \\
        Go & CWE-287: Improper Authentication & 60.25\% & CWE-208: Observable Timing Discrepancy & 0.00\% \\
        JavaScript & CWE-285: Improper Authorization & 22.57\% & CWE-116: Improper Encoding or Escaping of Output & 0.00\% \\
        Java   & CWE-681: Incorrect Conversion between Numeric Types & 40.63\% & CWE-1236: Improper Neutralization of Formula Elements in a CSV File & 0.00\% \\
        Python & CWE-522: Insufficiently Protected Credentials & 45.31\% & CWE-1021: Improper Restriction of Rendered UI Layers or Frames & 0.00\% \\
        \bottomrule
        \end{tabular}
    \end{adjustbox}
\end{table*}

\begin{table*}[t]
    \caption{Top 10 best and worst performing CWEs for GPT-4o and VulMaster in multilingual vulnerability repair}
    \label{tab:deep_analysis}
    \centering
    \small
    \begin{adjustbox}{max width=1.0\textwidth, center}
    
    \begin{tabular}{ccccccccc}
    \toprule
    \toprule
    & \multicolumn{4}{c}{GPT-4o}  & \multicolumn{4}{c}{VulMaster}   \\
    \midrule 
    Top 10   & CWE  & Name   & Accuracy & Count & CWE    & Name  & Accuracy & Count \\
\midrule 
 & {\color[HTML]{32CB00} 287} & {\color[HTML]{32CB00} Improper   Authentication}                                                                     & 72.09\%  & 129   & {\color[HTML]{32CB00} 287} & {\color[HTML]{32CB00} Improper Authentication}                                                                     & 72.09\%  & 129   \\
 & {\color[HTML]{32CB00} 306} & {\color[HTML]{32CB00} Missing Authentication for   Critical Function}                                                & 63.64\%  & 11    & 610                        & Externally Controlled Reference to a Resource in Another Sphere                                                    & 71.43\%  & 7     \\
 & {\color[HTML]{32CB00} 285} & {\color[HTML]{32CB00} Improper Authorization}                                                                        & 53.85\%  & 13    & {\color[HTML]{32CB00} 306} & {\color[HTML]{32CB00} Missing Authentication for Critical Function}  & 63.64\%  & 11    \\
 & 77  & Improper Neutralization of   Special Elements used in a Command ('Command Injection') & 47.37\%  & 38    & 521  & Weak Password Requirements & 54.55\%  & 11    \\
 & {\color[HTML]{32CB00} 362} & {\color[HTML]{32CB00} Concurrent Execution using   Shared Resource with Improper Synchronization ('Race Condition')} & 45.45\%  & 11    & {\color[HTML]{32CB00} 284} & {\color[HTML]{32CB00} Improper Access Control} & 52.17\%  & 23    \\
 & {\color[HTML]{32CB00} 79}  & {\color[HTML]{32CB00} Improper Neutralization of   Input During Web Page Generation ('Cross-site Scripting')} & 40.87\%  & 252   & 91 & XML Injection (aka Blind XPath Injection) & 46.15\%  & 13 \\
 & {\color[HTML]{CB0000} 359} & {\color[HTML]{CB0000} Exposure of Private Personal   Information to an Unauthorized Actor} & 40.00\%  & 5 & {\color[HTML]{32CB00} 285} & {\color[HTML]{32CB00} Improper Authorization} & 46.15\%  & 13    \\
 & 669 & Incorrect Resource Transfer Between Spheres & 40.00\%  & 5     & {\color[HTML]{32CB00} 362} & {\color[HTML]{32CB00} Concurrent Execution using Shared Resource with Improper Synchronization ('Race Condition')} & 45.45\%  & 11    \\
 & {\color[HTML]{CB0000} 532} & {\color[HTML]{CB0000} Insertion of Sensitive   Information into Log File} & 40.00\%  & 5 & 754 & Improper Check for Unusual or Exceptional Conditions & 45.45\%  & 11 \\
\multirow{-10}{*}{Best} & {\color[HTML]{32CB00} 284} & {\color[HTML]{32CB00} Improper Access Control}  & 39.13\%  & 23    & {\color[HTML]{32CB00} 79}  & {\color[HTML]{32CB00} Improper Neutralization of Input During Web Page Generation ('Cross-site Scripting')}        & 42.06\%  & 252   \\
\midrule & {\color[HTML]{CB0000} 401} & {\color[HTML]{CB0000} Missing Release of Memory after   Effective Lifetime} & 0.00\%   & 5 & {\color[HTML]{CB0000} 674} & {\color[HTML]{CB0000} Uncontrolled Recursion}  & 0.00\%   & 5 \\
 & {\color[HTML]{CB0000} 125} & {\color[HTML]{CB0000} Out-of-bounds Read} & 0.00\%   & 44    & {\color[HTML]{CB0000} 359} & {\color[HTML]{CB0000} Exposure of Private Personal Information to an Unauthorized Actor} & 0.00\%   & 5   \\
 & {\color[HTML]{CB0000} 835} & {\color[HTML]{CB0000} Loop with Unreachable Exit   Condition ('Infinite Loop')}  & 0.00\%   & 16    & {\color[HTML]{CB0000} 824} & {\color[HTML]{CB0000} Access of Uninitialized Pointer}                                                             & 0.00\%   & 5     \\
 & 134 & Use of Externally-Controlled   Format String  & 0.00\%   & 7 & 350  & Reliance on Reverse DNS Resolution for a Security-Critical Action   & 0.00\%   & 5     \\
 & 338 & Use of Cryptographically Weak   Pseudo-Random Number Generator (PRNG)  & 0.00\%   & 5     & {\color[HTML]{CB0000} 532} & {\color[HTML]{CB0000} Insertion of Sensitive Information into Log File}  & 0.00\%   & 5     \\
 & {\color[HTML]{CB0000} 824} & {\color[HTML]{CB0000} Access of Uninitialized Pointer} & 0.00\%   & 5     & 255 & Credentials Management Errors & 0.00\%   & 5     \\
 & {\color[HTML]{CB0000} 674} & {\color[HTML]{CB0000} Uncontrolled Recursion}  & 0.00\%   & 5     & {\color[HTML]{CB0000} 401} & {\color[HTML]{CB0000} Missing Release of Memory after Effective Lifetime}  & 0.00\%   & 5     \\
 & 369   & Divide By Zero  & 5.88\%   & 17    & 639  & Authorization Bypass Through User-Controlled Key & 0.00\%   & 6     \\
 & 617 & Reachable Assertion  & 5.88\%   & 17    & {\color[HTML]{CB0000} 125} & {\color[HTML]{CB0000} Out-of-bounds Read}  & 4.55\%   & 44 \\
\multirow{-10}{*}{Worst} & 415   & Double Free   & 7.69\%   & 13 & {\color[HTML]{CB0000} 835} & {\color[HTML]{CB0000} Loop with Unreachable Exit Condition ('Infinite Loop')}  & 6.25\%   & 16   \\
\bottomrule
\bottomrule
        \end{tabular}
    \end{adjustbox}
    \begin{tablenotes}
    \footnotesize
        \item[*] The CWE highlighted in green represents the best-performing CWE intersection between GPT-4o and VulMaster, while the CWE highlighted in red indicates the worst-performing CWE intersection between GPT-4o and VulMaster.
    \end{tablenotes}
\end{table*}

Table~\ref{tab:deep_analysis} presents the top 10 best and worst performing CWEs for GPT-4o with instruction-tuning and few-shot prompting strategies and VulMaster in the context of multilingual vulnerability repair.
To mitigate bias due to small sample sizes, we only consider CWEs with a count of at least 5.
As illustrated in the table, there are 6 CWEs that both GPT-4o and VulMaster perform best on, and 7 CWEs where both perform worst. 
Notably, there are 2 CWEs - CWE-359 (Exposure of Private Personal Information to an Unauthorized Actor) and CWE-532 (Insertion of Sensitive Information into Log File) - that are among the worst performing for VulMaster but are among the best performing for GPT-4o.
Moreover, although CWE-125 (Out-of-Bounds Read) and CWE-835 (Loop with Unreachable Exit Condition, i.e., 'Infinite Loop') are among the most frequently occurring CWEs where both GPT-4o and VulMaster perform poorly, GPT-4o failed to repair any instances of these vulnerabilities, whereas VulMaster successfully repaired at least some instances.
This suggests that while there is some overlap in performance between the two approaches, each also has unique strengths and weaknesses.
To further analyze these results, we draw upon CWE-1000~\cite{CWE1000} (Research Concepts), which is intended to support research into software weaknesses and their interdependencies, thereby providing a foundation for systematically identifying theoretical gaps within CWE. 
This view is primarily organized around behavioral abstractions and is explicitly designed to encompass all weaknesses defined in CWE.
By classifying each CWE into the CWE-1000 view, we observe that both GPT-4o and VulMaster perform well on CWE-284 (Improper Access Control) and CWE-707 (Improper Neutralization).
Conversely, both approaches perform poorly on CWE-664 (Improper Control of a Resource Through its Lifetime) and CWE-691 (Insufficient Control Flow Management).
The complete visualization results can be found in our replicated package.

\begin{tcolorbox}\textbf{RQ3 Summary:}
The instruction-tuned LLM (i.e., GPT-4o) with few-shot prompting, compared to AVR and PLM techniques, not only achieves the highest number of unique correct repairs but also completely avoids incorrect ones.
Furthermore, it performs relatively best on the top 25 most dangerous CWE-IDs, showcasing the potential of LLMs in repairing critical multilingual vulnerabilities.
\end{tcolorbox}

\begin{table}[t]
    \caption{Performance of LLMs on TypeScript vulnerability repair}
    \label{tab:typescript}
    \centering
    \tabcolsep=3.0mm
    \small
    \begin{adjustbox}{max width=1.0\textwidth, center}
    \begin{tabular}{lrrrrr}
        \toprule
        \toprule
        \textbf{Technique} & \textbf{EM with beam 1} & \textbf{BLEU-4} & \textbf{ROUGE-1} & \textbf{ROUGE-2} & \textbf{ROUGE-L} \\ 
        \midrule
        VulMaster & 5.88\%  & 0.3676 & 0.6877 & 0.4847 & 0.6741 \\
        GPT-4o    & 28.57\% & 0.7267 & 0.8453 & 0.7855 & 0.8439 \\
        \bottomrule
        \bottomrule
    \end{tabular}

    \end{adjustbox}
\end{table}

\subsection{RQ4: What is the generalization capability in repairing previously unseen vulnerabilities?}
\label{RQ4}
\noindent
\textbf{\emph{\underline{Approach.}}} 
Although current AVR methods and LLMs perform well on the REEF dataset, their ability to generalize to multilingual vulnerability repair remains uncertain.
To address this, we evaluate the leading AVR method, VulMaster, alongside the top-performing LLM, GPT-4o, utilizing instruction-tuning and few-shot prompting strategy on vulnerabilities in programming languages that have not been previously encountered.
Specifically, we extract vulnerabilities in the TypeScript programming language from the CVEfixes dataset, which provides both the vulnerabilities and their corresponding repairs at the function level. 
After retrieving the TypeScript data, we adhere to the data processing procedures outlined in Section \ref{subsec:dataset} and apply the same dataset division strategy.
This results in a collection of 593 TypeScript vulnerability instances, with 119 reserved for evaluation.

\noindent
\textbf{\emph{\underline{Results.}}} 
Table ~\ref{tab:typescript} presents the performance results of VulMaster and GPT-4o, employing instruction-tuning and few-shot prompting strategies for TypeScript vulnerability repair. 
As illustrated in the table, VulMaster achieves an EM score of 5.88\%, while GPT-4o attains a score of 28.57\%, which is 385.88\% higher than VulMaster. 
Furthermore, GPT-4o achieves a 28.71\% EM score on the REEF dataset, with its performance on TypeScript vulnerabilities surpassing this benchmark. 
In contrast, VulMaster, which performs at 28.94\% on the REEF dataset, experiences a significant decline in performance when applied to TypeScript vulnerabilities.
This suggests the robust generalization capabilities of GPT-4o compared to the limited generalization of VulMaster.
The superior performance of GPT-4o may be attributed to the strong comprehension abilities inherent in LLMs, whereas domain-specific AVR methods like VulMaster struggle to generalize to unseen data.
These findings underscore the potential of LLMs in multilingual vulnerability repair.

\begin{tcolorbox}\textbf{RQ4 Summary:}
On the unseen TypeScript vulnerabilities, GPT-4o with instruction-tuning and few-shot prompting strategies achieves an EM score of 28.57\%, outperforming VulMaster by 385.88\%.
These results demonstrate the strong generalization capability of LLMs in handling vulnerabilities in previously unseen programming languages.
\end{tcolorbox}

\section{Discussion}
\label{sec:discussion}
\subsection{Advanced prompting strategies}
Although we have investigated zero-shot prompting, few-shot prompting, and instruction-tuning strategies for multilingual vulnerability repair, several advanced prompting techniques, such as Chain-of-Thought (CoT)\cite{wei2022chain} remain underexplored and may yield different levels of performance.
CoT prompting guides LLMs to reason through problems step by step, significantly enhancing performance on reasoning-intensive tasks.
In this study, we take an initial step toward evaluating the effectiveness of CoT prompting in the context of multilingual vulnerability repair, with the goal of further enriching the scope and depth of our research.
To this end, we construct CoT prompts by extending the zero-shot prompt. Specifically, following prior work~\cite{yin2024thinkrepair}, we append the sentence "Let's think step by step." to the end of each instruction to form the CoT prompt.
We then assess the performance of both CoT prompting and instruction-tuning combined with CoT prompting across all open-source and closed-source LLMs evaluated in the preceding experiments.

\begin{table}[t]
    \caption{Performance of LLMs and CoT strategies on multilingual vulnerability repair}
    \label{tab:CoT}
    \centering
    \tabcolsep=3.0mm
    \small
    \begin{adjustbox}{max width=1.0\textwidth, center}
    \begin{tabular}{lrrrrr}
    \toprule
        \toprule
        \textbf{Technique} & \textbf{EM with beam 1} & \textbf{BLEU-4} & \textbf{ROUGE-1} & \textbf{ROUGE-2} & \textbf{ROUGE-L} \\ 
        \midrule
        \rowcolor{mygray}\multicolumn{6}{l}{\textbf{CoT prompting}} \\ 
        \midrule
        DeepSeek-Coder             & 0.14\%            & 0.1094 & 0.2796 & 0.1765 & 0.2746 \\
        Code Llama                 & 0.14\%            & 0.2109 & 0.5269 & 0.4218 & 0.5200 \\
        Llama 3                    & 0.09\%            & 0.0680 & 0.3200 & 0.1926 & 0.3181 \\
        GPT-3.5-Turbo              & 0.05\%            & 0.0576 & 0.4032 & 0.2090 & 0.4013 \\
        GPT-4o                     & 0.05\%            & 0.0755 & 0.3822 & 0.2008 & 0.3789 \\
        \midrule
        \rowcolor{mygray}\multicolumn{6}{l}{\textbf{Instruction-tuning   + CoT prompting}} \\ 
        \midrule
        DeepSeek-Coder             & 0.19\%            & 0.2464 & 0.4930 & 0.4625 & 0.4880 \\
        Code Llama                 & 1.82\%            & 0.3073 & 0.7381 & 0.6748 & 0.7308 \\
        Llama 3                    & 0.00\%            & 0.1854 & 0.7070 & 0.6236 & 0.7039 \\
        GPT-3.5-Turbo              & 5.79\%            & 0.7117 & 0.8435 & 0.7876 & 0.8386 \\
        GPT-4o                     & \textbf{7.38\%}   & \textbf{0.7707} & \textbf{0.8859} & \textbf{0.8397} & \textbf{0.8823} \\
        \bottomrule
        \bottomrule
    \end{tabular}

    \end{adjustbox}
\end{table}

Table~\ref{tab:CoT} presents the performance of various LLMs under different CoT prompting strategies on the multilingual vulnerability repair task.
Notably, GPT-4o, when combined with instruction-tuning and CoT prompting, outperforms all other open-source and closed-source LLMs across EM, BLEU, and ROUGE metrics, demonstrating its superior capability for this task—a finding that aligns with the conclusions drawn in RQ2.
Moreover, all closed-source LLMs show substantial performance gains when instruction-tuning is added to CoT prompting.
Specifically, when employing instruction-tuning and CoT prompting strategies, GPT-3.5-Turbo and GPT-4o achieve EM score of 5.79\% and 7.38\%, respectively.
In contrast, the same models using standard CoT prompting strategy yield significantly lower performance, with both GPT-3.5-Turbo and GPT-4o attaining only 0.05\% EM score.
In contrast, open-source LLMs exhibit mixed results under the same setting: Code Llama and DeepSeek-Coder benefit from instruction-tuning with CoT prompting, while Llama 3 shows a decline in performance.
This suggests that the impact of instruction-tuning on reasoning ability may vary significantly across LLMs.
Finally, even the best-performing LLM, GPT-4o with instruction tuning and CoT prompting, which achieves 7.38\% EM, falls significantly short of its performance with instruction tuning and few-shot prompting, which achieves 28.71\% EM (Table~\ref{tab:rq2}).
This observation suggests that CoT prompting may not be the most effective strategy for multilingual vulnerability repair.

\subsection{The effectiveness of larger LLMs}
In this study, we primarily evaluate open-source LLMs with 7B parameters, which demonstrate strong performance on the multilingual vulnerability repair task. 
Prior research~\cite{wei2022emergent, ganguli2022predictability} has shown that increasing model size can enhance the effectiveness of instruction tuning. 
However, other studies~\cite{mukherjee2023stack} have found that smaller, task-specialized models may outperform larger ones depending on the task. To explore whether this holds in our case, we extend our experiments to larger variants of the three open-source LLMs previously evaluated.
Specifically, we examine DeepSeek-Coder-33B-Instruct, Code-Llama-13B-Instruct, Code-Llama-34B-Instruct, Code-Llama-70B-Instruct, and LLaMA 3-70B-Instruct.
The training and inference settings are consistent with those described in Section~\ref{subsec:implementation}.
Based on the findings of RQ2, we adopt the most effective prompting strategy—instruction tuning combined with few-shot prompting—for all evaluations.

Table~\ref{tab:larger_models} presents the performance of the larger models on the multilingual vulnerability repair task. Overall, the results indicate substantial variation across different LLMs. 
Specifically, Llama 3 (70B) achieves the highest EM score among the larger models at 28.90\%, while DeepSeek-Coder (33B) attains a score of 20.31\%. In contrast, all models in the Code Llama series (13B, 34B, and 70B) perform poorly, with EM scores close to 0.00\%, underperforming their smaller counterparts.
Compared to the results shown in Table~\ref{tab:rq2}, Llama 3 (70B) outperforms Llama 3 (8B) by 57.92\%, with the latter achieving an EM score of 18.30\%.
However, DeepSeek-Coder (6.7B) achieves a higher EM score of 20.96\%, outperforming its 33B counterpart by 3.20\%.
Similarly, Code Llama (7B) reaches an EM score of 18.25\%, while all larger variants in the same series almost entirely fail to repair any vulnerabilities.
These results suggest that although larger open-source LLMs have the potential to enhance performance on the multilingual vulnerability repair task, improvements are not guaranteed; in some cases, smaller and more efficient models may yield better results.

\begin{table}[t]
    \caption{Performance of larger LLMs on multilingual vulnerability repair}
    \label{tab:larger_models}
    \centering
    \tabcolsep=3.0mm
    \small
    \begin{adjustbox}{max width=1.0\textwidth, center}
    \begin{tabular}{lrrrrr}
        \toprule
        \toprule
        \textbf{Technique} & \textbf{EM with beam 1} & \textbf{BLEU-4} & \textbf{ROUGE-1} & \textbf{ROUGE-2} & \textbf{ROUGE-L} \\ 
        \midrule
        \rowcolor{mygray}\multicolumn{6}{l}{\textbf{Instruction-tuning   + Few-shot prompting}} \\ 
        \midrule
        DeepSeek-Coder (33B) & 20.31\% & \textbf{0.6718} & 0.7600  & 0.7235 & 0.7556  \\
        Code Llama (13B)     & 0.00\% &0.0211  & 0.0475 & 0.0215 & 0.0450   \\
        Code Llama (34B)     & 0.00\% & 0.0238  & 0.0507 & 0.0244 & 0.0479 \\
        Code Llama (70B)     & 0.47\% & 0.1836 & 0.3925 & 0.2637 & 0.3781 \\
        Llama 3 (70B)        & \textbf{28.90\%} & 0.5067  & \textbf{0.8644} & \textbf{0.8048} & \textbf{0.8627} \\
        \bottomrule
        \bottomrule
    \end{tabular}
    \end{adjustbox}
\end{table}

\subsection{Explanations for the low fix rate on multilingual vulnerability}
The instruction-tuning GPT-4o with few-shot prompting strategy showed varying performance across programming languages, with the best results for Go and the worst for C/C++. 
To understand the reasons for incorrect fixes and improve future LLM-based approaches, we manually analyzed GPT-4o's results. 
We randomly selected 200 vulnerable-fixed code pairs from the test set and examined the incorrect fixes generated for both Go and C/C++.
Specifically, the first author completed the manual coding analysis by referring to the work of \citet{zhou2024out}.
Five reasons were classified: error localization, multiple chunks errors, logical errors, context format errors, and semantic equivalence.

\begin{table}[b]
\caption{Manual check result for GPT-4o on Go and C/C++}
    \label{tab:5_manual}
    \begin{tabular}{lrr}
    \toprule
    Language & Go & C/C++ \\
    \midrule
    Error localization & 49 & 43 \\
    Multiple chunk errors & 12 & 9 \\
    Logical errors & 29 & 25 \\
    Context format errors & 5 & 8 \\
    Semantic equivalence & 5 & 15 \\
    Total & 100 & 100 \\
    \bottomrule
    \end{tabular}
\end{table}

\begin{figure*}[t]
    \centering
    \includegraphics[width=.8\linewidth]{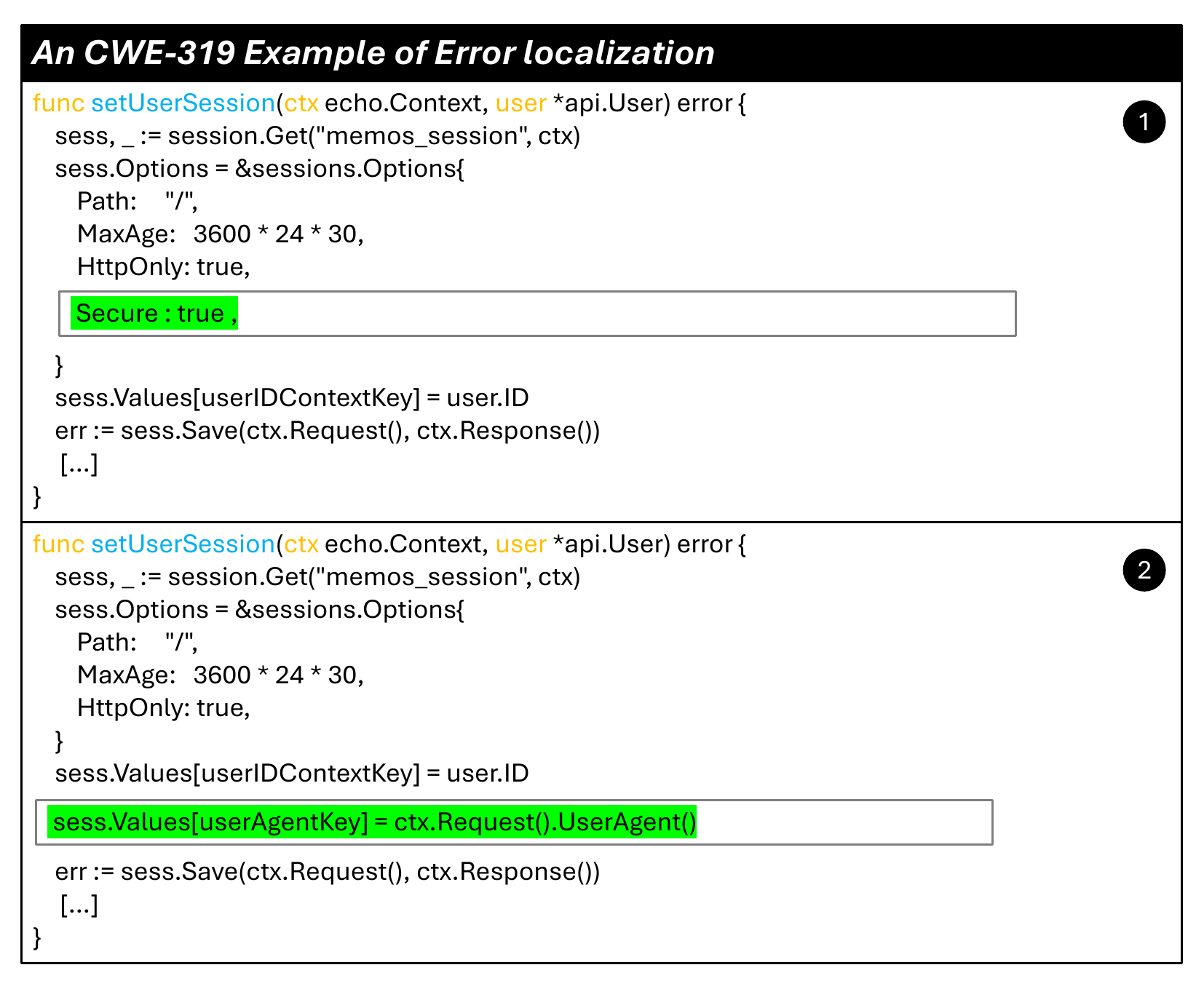}
    \caption{A CWE-319 example of error localization in Go language. Code 1 is the repair code and Code 2 is generated by GPT-4o. The code within the brackets highlights the modifications, with additions shown on a green background and deletions on a red background. }
    \label{fig:5_localization_error}
\end{figure*}

Table~\ref{tab:5_manual} presents the frequency of five main reasons for incorrect fixes.
As shown in the table, a significant portion of the errors in the GPT-4o results resulted from localization issues (49\% and 43\% for Go and C/C++, separately).
Additionally, many instances involved multiple or incorrect modifications to the original code.
The remaining fixes could be classified as correct vulnerability repairs, with the following observations:
GPT-4o did not fully learn the specific data format we employed, resulting in multiple token-level differences in the output.
Some of the fixes achieved semantic equivalence, which indicates a partial understanding of the task.
In terms of semantic equivalence, GPT-4o performed relatively better in C/C++, with 23 out of 100 examples achieving this, while Go performed worse in this aspect. 
However, the overall accuracy of GPT-4o on Go was already higher, so the slight performance gap is understandable. (Accuracy: Go – 37.72\%, C – 15.08\%, C++ – 13.82\%).

\begin{figure*}[]
    \centering
    \includegraphics[width=.8\linewidth]{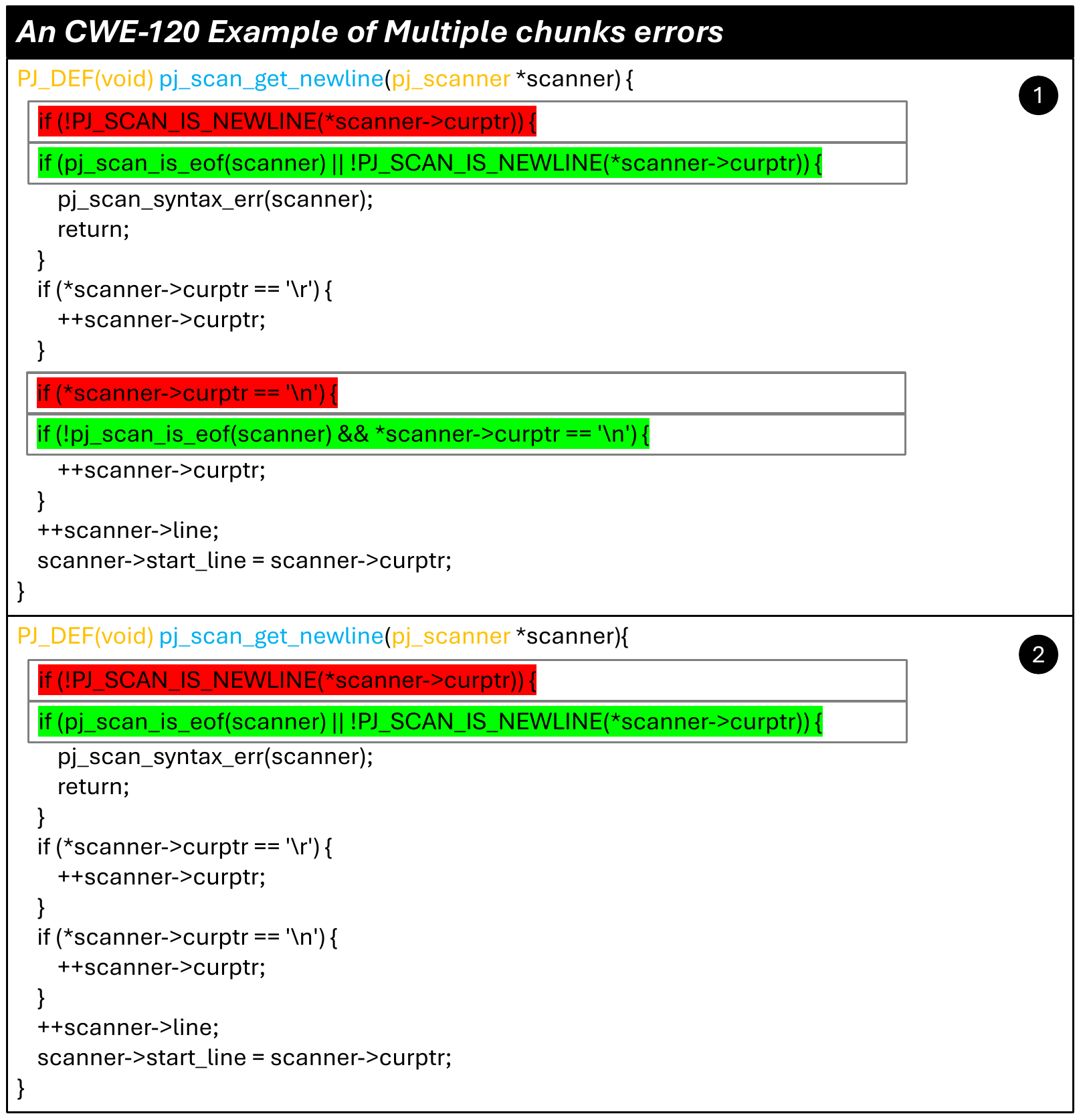}
    \caption{A CWE-120 example of multiple chunks errors in C language. Code 1 is the repair code and Code 2 is generated by GPT-4o. The code within the brackets highlights the modifications, with additions shown on a green background and deletions on a red background. }
    \label{fig:5_multi_chunks}
\end{figure*}

\begin{figure*}[]
    \centering
    \includegraphics[width=.8\linewidth]{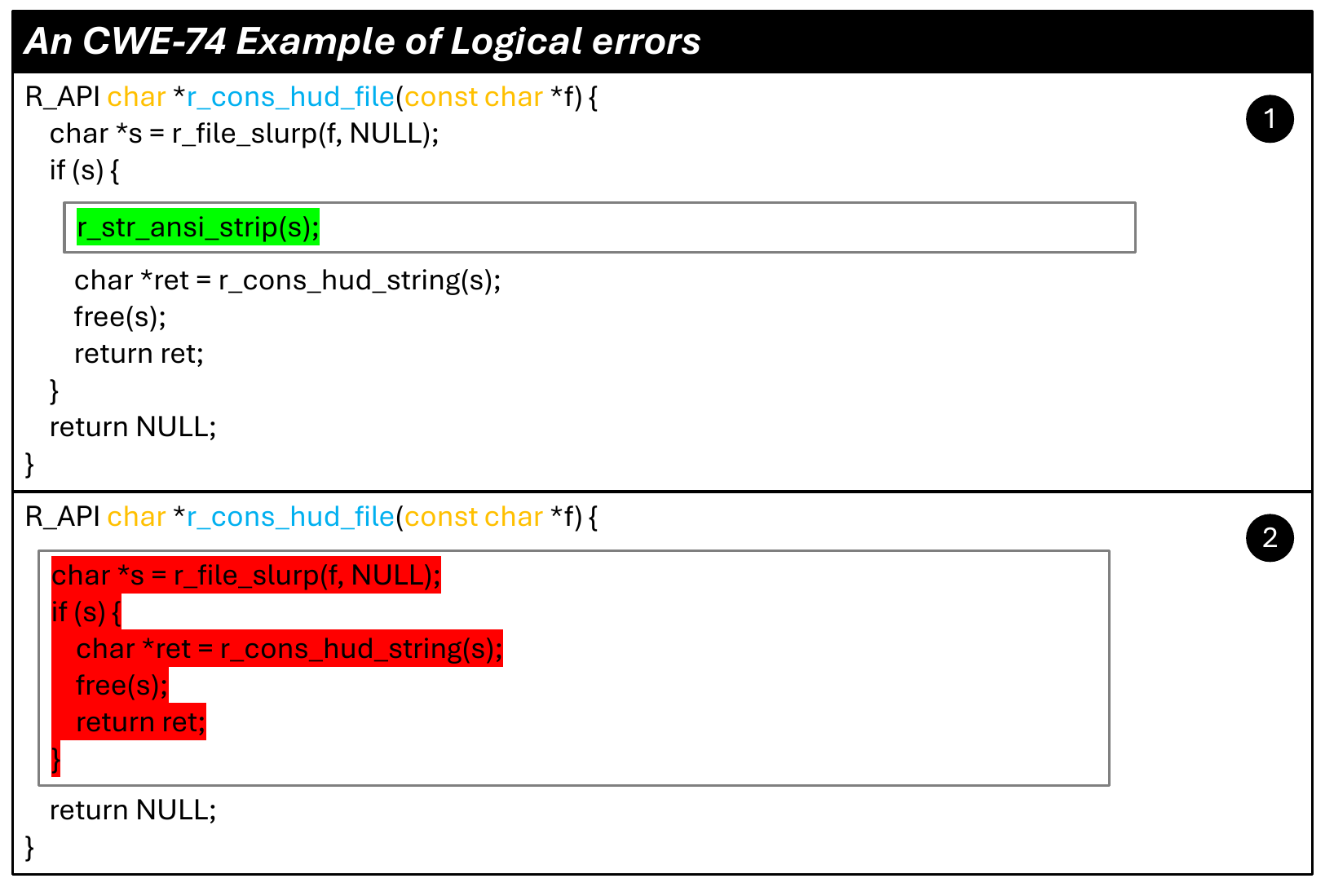}
    \caption{A CWE-74 example of logical error in Go language. Code 1 is the repair code and Code 2 is generated by GPT-4o. The code within the brackets highlights the modifications, with additions shown on a green background and deletions on a red background. }
    \label{fig:5_logical_error}
\end{figure*}

We now selected several representative examples to illustrate the reasons for the failure.
Figure~\ref{fig:5_localization_error} exhibits a CWE-319 (i.e., Cleartext Transmission of Sensitive Information) example of localization error in the Go language.
In Code 1, the added code ensures that session cookies are transmitted only over HTTPS connections, preventing the transmission of sensitive session information over insecure HTTP.
This configuration effectively reduces the risk of session cookies being intercepted and prevents sensitive information from being exposed in unencrypted transmissions.
However, in Code 2, GPT-4o failed to identify this vulnerability, making modifications in unrelated areas instead.
Figure~\ref{fig:5_multi_chunks} shows a CWE-120 (i.e., Buffer Copy without Checking Size of Input) example of multiple chunks errors in C language.
In Code 1, the changed code addresses this issue by adding an EOF check in the condition statements. 
This ensures that the pointer is valid and within the bounds of the buffer before accessing its value, preventing potential buffer overflows and fixing CWE-120.
In contrast, in Code 2, GPT-4o failed to identify all the vulnerable statements, leaving the vulnerability unresolved.
Figure~\ref{fig:5_logical_error} shows a CWE-74 (i.e., Improper Neutralization of Special Elements in Output Used by a Downstream Component) example of logical error in C language.
In Code 1, the vulnerability is potentially mitigated by the addition of the r\_str\_ansi\_strip(s) function call, which removes ANSI control characters from the input string `s' and thereby reduces the risk of injection attacks exploiting special characters.
However, in Code 2, GPT-4o failed to address this vulnerability effectively and instead removed all related code, which is not a valid solution.
Based on our manual analysis, future research could investigate the use of multi-agent LLMs to tackle localization tasks in vulnerability repair, which could further enhance the performance of LLMs in multilingual vulnerability repair.

\subsection{The security of LLMs for vulnerability repair}
Although LLMs have demonstrated remarkable capabilities in multilingual vulnerability repair, the use of LLM-generated patches poses significant security risks.
First, over-reliance on LLM-generated patches without adequate human oversight may inadvertently introduce new vulnerabilities.
Although these models can identify and address known issues, they may also produce patches that only superficially resolve the problem while overlooking deeper security implications.
\citet{fu2023security} analyzed code snippets generated by GitHub Copilot and found that developers face a high risk of introducing security weaknesses when using Copilot or other AI-based code generation tools, regardless of the programming language.
Therefore, implementing appropriate security checks is essential.
Second, the internal architectures, training data, and inference mechanisms of closed-source LLMs remain opaque.
This lack of transparency raises concerns about the potential presence of backdoors and inherent model biases, which are difficult for users to detect or mitigate. 
In security-critical contexts, such opacity severely undermines trust, accountability, and verifiability. Moreover, the use of closed-source LLMs raises substantial concerns about data privacy and security, as users have limited control over how their inputs are processed and stored.
Third, a recent study~\cite{wang2024your} has revealed that current LLMs often overlook important security considerations during both code generation and repair.
As a result, the output code may contain latent vulnerabilities or subtle logic flaws that compromise system security.
In summary, while LLMs hold great promise in automated vulnerability repair, their current limitations—especially in closed-source settings—necessitate caution.
Without rigorous validation and human-in-the-loop verification, LLM-generated patches may not only fail to address the original issue but may also introduce new security risks.

\subsection{The promise of LLMs on multilingual vulnerability repair}
Results from the comparative analysis of RQ1 through RQ4 show that LLMs with appropriate strategies compete effectively with AVR techniques and PLM approaches in multilingual repair tasks.
For instance, GPT-4o achieves a 28.71\% EM rate, comparable to VulMaster's 28.94\% EM.
Specifically, by combining instruction-tuning with BM25-based few-shot prompting, LLMs can generate high-quality repairs for inputted vulnerable functions without needing side information (e.g., CWE details) or specialized pre-processing (e.g., special tokens or program analysis). 
Due to the language-agnostic capabilities of LLMs, they can also achieve competitive results with limited computational resources, even without instruction-tuning.
For instance, GPT-4o with sole using few-shot prompting achieves a 26.89\% EM, significantly outperforming many conventional AVR techniques and PLM approaches.
Furthermore, GPT-4o, when equipped with instruction tuning and few-shot prompting, demonstrates strong generalization to unseen TypeScript vulnerabilities, whereas the leading AVR technique, VulMaster, exhibits only limited generalization.
These findings show that LLMs have promising capabilities for multilingual vulnerability repair tasks and substantial untapped potential.
We identify five future directions for enhancing LLM-based multilingual vulnerability repair.

First, future work could explore additional prompting strategies beyond those examined in this study, which focused solely on zero-shot prompting, few-shot prompting, instruction tuning, and CoT prompting.
Advanced techniques such as Tree-of-Thought (ToT) prompting~\cite{yao2023tree} and iterative refinement based on feedback~\cite{kulsum2024case} merit further investigation. 
Specifically, researchers could leverage static analysis security tools to evaluate the code repaired by AVR techniques and incorporate the resulting feedback to guide and enhance LLM outputs.
Moreover, the current few-shot prompting approach uses an IR-based BM25 method for selecting examples, which relies on syntactic similarity.
Future work could explore alternative example selection methods to ensure a more diverse set of examples in few-shot prompts, thereby improving the model’s ability to generalize to new inputs.
Second, the scale of the dataset we use remains relatively small, which limits the performance of instruction-tuned LLMs to some extent. 
Future research could expand multilingual vulnerability benchmarks by leveraging the framework of multilingual vulnerability data collection~\cite{wang2023reef}.
Third, the most recent study ~\cite{sheng2025llms} indicates that successful vulnerability repair in production requires that the patched code pass all existing tests, prevent recurrence of the vulnerability, and not introduce new security issues.
However, generally existing available datasets are not executable for verification, making it difficult to verify these criteria.
Future work should evaluate AVR techniques on dynamically verifiable datasets to better ensure the reliability and applicability of the results in real-world settings.
Moreover, current LLMs for multilingual vulnerability repair only use a single modality (i.e., source code), which may limit the effectiveness of LLMs on some specific vulnerability type.
For example, recent 
studies~\cite{cao2022mvd, shiri2024systematic} indicate that leveraging control-flow and data-flow graph is beneficial for detecting the memory-related vulnerability.
Therefore, we should consider incorporating additional data modality like graph-structural data to improve the performance of LLMs on particular vulnerabilities.
Last, recent studies~\cite{honarvar2025evaluating, chen2024nlperturbator}have underscored the importance of robustness testing for LLMs in security-critical tasks. 
\citet{chen2024nlperturbator} demonstrated that real-world variations in natural language prompts can substantially degrade the performance of code LLMs.
Similarly, \citet{honarvar2025evaluating} emphasized that LLM-based code generation suffers from generalization gaps across semantically related tasks.
These findings highlight the need for further investigation into the robustness of AVR techniques against minor code modifications or adversarial inputs.
Given that LLMs are highly sensitive to input variations, where even slight differences in semantically equivalent programs can lead to repair failures, comprehensive robustness testing is essential before deploying AVR techniques in practical settings.

\section{Threats to Validity}
\label{sec:threats}
\textit{External Threats} mainly lie in the used multilingual vulnerability dataset and the studied automated vulnerability repair approaches.
Since this work relied on the REEF dataset, which encompasses seven popular programming languages, our findings may not generalize to other languages. 
For future work, we plan to expand our collection of vulnerabilities across a broader range of programming languages using the REEF framework.
To mitigate the second threat, we conducted a systematic literature review, ensuring that our selected automated vulnerability repair approaches are state-of-the-art and representative.
Regarding the LLM selection, we consider both advanced open-source and closed-source LLMs to ensure diversity.

\textit{Internal Threat} arises from our implementation of the studied automated repair approaches. 
To mitigate this threat, we implemented all existing approaches using the replication packages provided in their respective papers and two authors carefully reviewed the source code.
For the studied LLMs, we used publicly accessible models (i.e., DeepSeek-Coder, Code Llama, and Llama3) from Hugging Face or invoked their APIs (i.e., GPT-3.5-Turbo and GPT-4o) as per official instructions. 
Another potential threat to validity stems from the non-deterministic nature of LLMs, which may influence the experimental results and findings.
To mitigate this issue and support future reproducibility, we have explicitly documented the configuration settings of all LLMs used in this study and publicly released the fully reproducible replication package ~\cite{AVR2025}.

\textit{Construct Threats} primarily stem from the construction of vulnerable function pairs and metrics that evaluate the automated vulnerability approaches. 
We relied on the tool namely Tree-sitter to parse the commit data and collect function pairs, which means there is a possibility of incorrect function pair matching.
To mitigate this threat, we performed a sanity check on a group of randomly selected samples to ensure the robustness of the tool. 
For our evaluation metrics, we used established measures in the vulnerability domain, including Exact Match, BLEU, and ROUGE scores.

\section{Conclusion}
\label{sec:conclusion}
This work presents a large-scale empirical study to examine the effectiveness of the existing software vulnerability repair approaches (learning-based approaches and pre-trained language models) and LLMs in the context of multilingual vulnerabilities across seven programming languages.
Our key findings show that GPT-4o with instruction tuning and few-shot prompting is both the best-performing LLM and competitive with VulMaster, the leading AVR approach, in multilingual vulnerability repair.
Furthermore, instruction-tuned GPT-4o with few-shot prompting outperforms VulMaster in repairing unique vulnerabilities and is more effective at addressing the most dangerous vulnerabilities.
Our manual analysis reveals that error localization is the main factor preventing instruction-tuned GPT-4o from successfully repairing vulnerabilities.

Meanwhile, our work opens up several promising future research directions, including:
exploring advanced prompting strategies, expanding the scale of multilingual vulnerability datasets, evaluating AVR techniques on dynamically verifiable benchmarks, incorporating multimodal approaches to address specific vulnerability types across diverse programming languages, and performing comprehensive robustness testing prior to real-world deployment.

\section{Data Availability}
\label{sec:data_availability}
We released all the experimental data and source code on the project homepage for replication, future research, and practical use~\cite{AVR2025}.

\begin{acks}
This work was supported by National Natural Science Foundation of China (Grant Nos. 62322208, 12411530122), JSPS for the KAKENHI grants (JP21H04877, JP22K18630), Bilateral Program grant JPJSBP120239929, Japan Science and Technology Agency (JST) as part of Adopting Sustainable Partnerships for Innovative Research Ecosystem (ASPIRE), Grant Nos JPMJAP2415, and the Inamori Research Institute for Science for supporting Yasutaka Kamei via the InaRIS Fellowship. 
\end{acks}

\balance
\bibliographystyle{ACM-Reference-Format}
\bibliography{reference}

\end{document}